\documentclass[aps,pra,twocolumn,showpacs,superscriptaddress]{revtex4}
\usepackage{graphicx}
\usepackage{psfrag}
\usepackage{amsmath}
\usepackage{amssymb}
\usepackage{bm}
\newcount\minute
\newcount\hour
\def\currenttime{%
    \minute\time
    \hour\minute
    \divide\hour60
    \the\hour:\multiply\hour60\advance\minute-\hour\the\minute}
\renewcommand{\vec}[1]{\boldsymbol{#1}}
\renewcommand{\tensor}[1]{\overset{\,\leftrightarrow\!}{#1}}
\newcommand{\Nabla}{\vec{\nabla}}
\newcommand{\Lnabla}{\overset{\leftarrow}{\Nabla}}
\newcommand{\bra}[1]{\left\langle #1 \right |}
\newcommand{\ket}[1]{\left| #1 \right\rangle}
\newcommand{\EW}[1]{\langle #1 \rangle}

\newcommand{\D}{\mathrm{d}}
\renewcommand{\Im}{{\rm Im}\,}
\renewcommand{\Re}{{\rm Re}\,}
\newcommand{\uv}[1]{\mathbf{e}_{#1}}

\newcommand{\fo}[1]{\underline{\hat{\mathbf{#1}}}}
\newcommand{\fon}[1]{\underline{\mathbf{#1}}}
\newcommand{\ot}{}         
\begin{document}
\title{Unified approach to QED in arbitrary linear media}
\author{Christian Raabe}
\affiliation{Theoretisch-Physikalisches Institut,
Friedrich-Schiller-Universit\"at Jena, Max-Wien-Platz 1, D-07743 Jena,
Germany}
\author{Stefan Scheel}
\affiliation{Quantum Optics and Laser Science, Blackett Laboratory,
Imperial College London, Prince Consort Road, London SW7 2BW, United
Kingdom}
\author{Dirk-Gunnar Welsch}
\affiliation{Theoretisch-Physikalisches Institut,
Friedrich-Schiller-Universit\"at Jena, Max-Wien-Platz 1, D-07743 Jena,
Germany}
\date{\today, \currenttime}


\begin{abstract}
We give a unified approach to macroscopic
QED in arbitrary linearly responding media,
based on the quite general, nonlocal form of
the conductivity tensor as
it can be introduced within the
framework of linear response theory, and appropriately chosen
sets of bosonic variables.
The formalism generalizes the quantization schemes
that have been developed previously
for diverse classes of linear media.
In particular, it turns out that the scheme developed for locally
responding linear magnetodielectric media can be recovered
from the general scheme as a limiting case
for weakly spatially dispersive media.
With regard to practical applications, we furthermore address
the dielectric approximation for the conductivity tensor and
the surface impedance method for the calculation of the Green tensor
of the macroscopic Maxwell equations,
the two central quantities of the theory.
\end{abstract}
\maketitle
\section{Introduction}
\label{sec1}

In both classical and quantum electrodynamics, it is often advisable
to divide, at least notionally, the matter that interacts with the
electromagnetic field into a part that plays the
role of a passive background and a remainder,
active part that needs to be considered in more detail.
By means of suitable coarse-graining and
averaging procedures, this leads to
the well-known framework of Maxwell's phenomenological equations,
where the background---the medium---is treated as a continu\-um
and, quite frequently,
by the methods of linear response theory.
From this perspective, the characterization
of the medium is reduced to the prescription of suitable constitutive
relations, i.e., appropriate response functions or susceptibilities.

Depending on the specific kinds of media under consideration,
it is under many circumstances sufficiently
accurate to work with spatially
local response functions, taking into account only
(temporal) dispersion and absorption
in accordance with causality.
For conducting and semiconducting media
(not to mention plasmas) as well as superconducting materials,
however, the spatially local description
can be inadequate due to the existence of almost
freely movable charge carriers
(conduction electrons, excitons, Cooper pairs)
in such media. Hence, if one is not willing to restrict one's attention
to a crude spatial resolution and/or specific frequency windows, spatial
dispersion, i.e., the spatially non-local character of the medium
response, generally cannot be disregarded for such media.
Electrodynamics problems with the inclusion of spatial dispersion
have been considered by various authors in different ways,
both on the classical and quantum levels; for classical approaches,
see, e.g., Refs.~\cite{BirmanJL091972,MaradudinAA031973,AgarwalGS101972,
AgarwalGS081974,AgarwalGS021975,GinzburgED,MelroseBook},
for quantum ones see, e.g., Refs.~\cite{SavastaS032002,SavastaS082002}.

A scheme that takes spatial dispersion into account
along with dispersion and absorption in sufficiently general terms
can also be regarded as an important step
towards a satisfactory (quantum) electrodynamics
of moving media, which is very much lacking at present.
The reason is that a medium, even if it can
be assumed to respond spatially locally
when it is at rest, will in general appear as
responding non-locally when it is in motion.
Given that the polarization of
typical Drude--Lorentz-type dielectrics responds to the electric field with
a characteristic memory time of the order of $10^{-9}\cdots 10^{-7}$s
\cite{Jackson3rd}, already moderate (i.e., non-relativistic) velocities
may lead to the appearance of noticeable spatial non-localities.
For example, sonoluminescence experiments show that the collapse of a
bubble with a typical initial radius of $10\cdots 50\mu$m to a final
radius of around $1\mu$m occurs on a time scale
similar to the characteristic memory time of the response
of the surrounding fluid \cite{SonoluminescenceReview2}.

The study of the quantized electromagnetic field in
spatially non-locally responding media
and the prospect of elaborating a quantum theory of light in moving media
will also open up new ways of investigating quantum effects related to
the recently proposed `optical black hole'
\cite{LeonhardtU011999,LeonhardtU012000}. So far, the theory has
concentrated on purely geometrical optics with some progress being made
towards a (scalar) wave-optical description,
but a consistent linear-response approach is still lacking.

The paper is organized as follows. In Sec.~\ref{sec2} we introduce the
basic concepts of field quantization in
arbitrary linearly responding media,
with special emphasis on spatially dispersive media.
This serves as the basis for a detailed study of
possible choices of appropriate dynamical variables
in Sec.~\ref{sec3}. We
then proceed to show in Sec.~\ref{sec4} how
previously introduced quantization schemes for
diverse classes of media
can be obtained
as special cases from the general quantization
scheme developed in Sec.~\ref{sec2}.
In addition to the general formalism, some knowledge of
the structure of the Green tensor for spatially dispersive media is
needed when performing explicit calculations. This problem is addressed
in Sec.~\ref{sec5}, where it is described (in general and
by an example) how the surface impedance method may be applied in
this context, on the basis of the dielectric approximation.
Some concluding remarks are given in Sec.~\ref{sec6}.
%

\section{Quantization scheme}
\label{sec2}

The effect of any linear, dispersing and absorbing medium on the
electromagnetic field can be described,
within the framework of linear response theory, by the relation
\begin{equation}
\label{eq8-1}
\fon{j}(\mathbf{r},\omega)=\int
\D^3r'\,\tensor{Q}(\mathbf{r},\mathbf{r}',\omega)
\cdot\fon{E}(\mathbf{r'},\omega)
+\fon{j}_{\mathrm N}(\mathbf{r},\omega),
\end{equation}
where $\fon{j}(\mathbf{r},\omega)$ and $\fon{E}(\mathbf{r},\omega)$,
respectively, are the (linearly responding)
current density and the electric field in the
frequency domain, $\tensor{Q}(\mathbf{r},\mathbf{r}',\omega)$
is the complex conductivity tensor in the frequency domain
\cite{KuboNonEqStat,MelroseBook},
and $\fon{j}_{\mathrm N}(\mathbf{r},\omega)$ is
a Langevin noise source.
According to the Onsager reciprocity theorem
\cite{KuboNonEqStat,MelroseBook}, the
conductivity tensor
should be
reciprocal,
$Q_{ij}(\mathbf{r},\mathbf{r}',\omega)$ $\!=$
$\!Q_{ji}(\mathbf{r}',\mathbf{r},\omega)$.
Except for a translationally invariant (bulk) medium, the
spatial arguments $\mathbf{r}$ and $\mathbf{r}'$
of $\tensor{Q}(\mathbf{r},\mathbf{r}',\omega)$
must be kept as two separate variables in general.
We assume that, for chosen $\omega$,
$\tensor{Q}(\mathbf{r},\mathbf{r}',\omega)$
is the integral kernel of a reasonably well-behaved
(integral) operator acting on vector functions in
position space. In particular,
we assume that $\tensor{Q}(\mathbf{r},\mathbf{r}',\omega)$ tends
(sufficiently rapidly) to zero for $|\mathbf{r-r'}|\to\infty$
and has no strong (i.e., non-integrable)
singularities (specifically, for $\mathbf{r'}\to\mathbf{r}$).
To allow for the spatially non-dispersive limit,
$\delta$-functions and their derivatives
must be permitted so that
$\tensor{Q}(\mathbf{r},\mathbf{r}',\omega)$ may become
a (qua\-\mbox{si-)l}o\-cal integral kernel.
In the remainder of the paper, we will use the
superscripts~${}^\mathsf{T}$ and~${}^{+}$
to indicate transposition and Hermitian conjugation
with respect to tensor indices.
Since the spatial arguments are not switched by these operations,
the operator associated with an integral kernel
$\tensor{A}(\mathbf{r},\mathbf{r}')$
is Hermitian if $\tensor{A}(\mathbf{r},\mathbf{r}')$
$\!=$ $\!\tensor{A}{^{+}}(\mathbf{r}',\mathbf{r})$.
In particular, an operator associated with a
real kernel is Hermitian if it has the reciprocity property
$\tensor{A}(\mathbf{r},\mathbf{r}')$
$\!=$ $\!\tensor{A}{^{\mathsf{T}}}(\mathbf{r}',\mathbf{r})$.
The decomposition
$\tensor{Q}(\mathbf{r},\mathbf{r}',\omega)$ $\!=$
$\!\Re\tensor{Q}(\mathbf{r},\mathbf{r}',\omega)$ $\!+$
$\!i\,\Im\tensor{Q}(\mathbf{r},\mathbf{r}',\omega)$
of the conductivity tensor is therefore identical
with the decomposition of the associated operator
into a Hermitian and an anti-Hermitian part,
\begin{align}
\label{eq2}
\tensor{\sigma}(\mathbf{r},\mathbf{r}',\omega)
&\equiv \Re\tensor{Q}(\mathbf{r},\mathbf{r}',\omega)
\nonumber\\
&={\textstyle\frac{1}{2}}
\bigl[\tensor{Q}(\mathbf{r},\mathbf{r}',\omega)+
\tensor{Q}{^{+}}(\mathbf{r}',\mathbf{r},\omega)\bigr],
\end{align}
\begin{align}
\label{eq3}
\tensor{\tau}(\mathbf{r},\mathbf{r}',\omega)
&\equiv
\Im\tensor{Q}(\mathbf{r},\mathbf{r}',\omega)
\nonumber\\
&={\textstyle\frac{1}{2i}}\bigl[\tensor{Q}(\mathbf{r},\mathbf{r}',\omega)
-\tensor{Q}{^{+}}(\mathbf{r}',\mathbf{r},\omega)\bigr].
\end{align}
Since $\tensor{\sigma}(\mathbf{r,r'},\omega)$
is associated with the dissipation of electromagnetic
energy (see, e.g., Refs.~\cite{KuboNonEqStat,MelroseBook}),
the operator associated with the integral kernel
$\tensor{\sigma}(\mathbf{r},\mathbf{r}',\omega)$ is,
for real $\omega$,
a positive definite operator in the case of
absorbing media considered throughout this paper.

The conductivity tensor $\tensor{Q}(\mathbf{r},\mathbf{r}',\omega)$
is the temporal Fourier transform of a response
function $\tensor{\tilde{Q}}(\mathbf{r,r'},t)$ in the time domain,
\begin{equation}
\label{eq8-1-1}
\tensor{Q}(\mathbf{r,r'},\omega)
= \int \D t \, e^{i\omega t}
\tensor{\tilde{Q}}(\mathbf{r,r'},t),
\end{equation}
which satisfies causality conditions of the type
\begin{equation}
\label{NL1c}
\tensor{\tilde{Q}}(\mathbf{r,r'},t)
= 0
\quad\mathrm{if}\quad
t-\cos\eta\,|\mathbf{r-r'}|/c < 0
\end{equation}
for chosen $\mathbf{r}$ and $\mathbf{r'}$ and
arbitrary directional cosines
$\cos\eta$ ($0\!\le\!\cos\eta\!\le\!1$).
In particular, for $\cos\eta$ $\!=$ $\!0$, one finds from arguments
\cite{LanLifStat,NussenzweigBook,KuboNonEqStat}
similar to those for the case of spatially locally
responding media that, for chosen $\mathbf{r}$ and $\mathbf{r'}$,
$\tensor{Q}(\mathbf{r},\mathbf{r}',\omega)$
is analytic in the upper complex $\omega$ half-plane,
fulfills Kramers--Kronig (Hilbert transform) relations, and satisfies
the Schwarz reflection principle
$\tensor{Q}{^{\ast}}(\mathbf{r,r'},\omega)\!=\!
\tensor{Q}(\mathbf{r,r'},-\omega^{\ast})$.
Other values of $\cos\eta$ could
obviously provide more stringent (spatio-temporal) conditions
(see also Ref.~\cite{MelroseDB011977}),
which are, however, not required here.

Let us identify the current density
that enters the macroscopic Maxwell equations
in the frequency domain
with $\fon{\mathbf{j}}(\mathbf{r},\omega)$
as specified in Eq.~(\ref{eq8-1}).
In this case, the medium-assisted
electric field in the frequency
domain satisfies the integro-differential equation
\begin{multline}
\label{eq8-2}
\Nabla\times\Nabla\times\fon{E}(\mathbf{r},\omega)
-\frac{\omega^2}{c^2}\,\fon{E}(\mathbf{r},\omega)
\\
-i\mu_{0}\omega
\int\D^3r'\,\tensor{Q}(\mathbf{r,r'},\omega)
\cdot\fon{E}(\mathbf{r'},\omega)
=i\mu_{0}\omega \fon{j}_{\mathrm N}(\mathbf{r},\omega),
\end{multline}
whose unique solution is
\begin{equation}
\label{NL7}
\fon{E}(\mathbf{r},\omega)=i\mu_{0}\omega\int \D^3r'\,
\tensor{G}(\mathbf{r,r'},\omega)
\cdot\fon{j}_{\mathrm N}(\mathbf{r'},\omega),
\end{equation}
with $\tensor{G}(\mathbf{r},\mathbf{r}',\omega)$ being
the (retarded) Green tensor. It satisfies Eq.~(\ref{eq8-2})
with the (tensorial) $\delta$-function source,
\begin{multline}
\label{eq8-3}
\Nabla\times\Nabla\times
\tensor{G}(\mathbf{r,s},\omega)
-\frac{\omega^2}{c^2}\,\tensor{G}(\mathbf{r,s},\omega)
\\
-i\mu_{0}\omega\int
\D^3r'\,\tensor{Q}(\mathbf{r,r'},\omega)
\cdot
\tensor{G}(\mathbf{r',s},\omega)
=
\tensor{I}\delta(\mathbf{r-s}),
\end{multline}
together with the boundary condition
at infinity, and has all the attributes
of a (Fourier transformed) causal response function
just as $\tensor{Q}(\mathbf{r},\mathbf{r}',\omega)$
has them. In particular, it is analytic in
the upper $\omega$ half-plane and the
Schwarz reflection principle
$\tensor{G}{^{\ast}}(\mathbf{r},\mathbf{r}',\omega)$
$\!=\tensor{G}(\mathbf{r},\mathbf{r}',-\omega^{\ast})$ is valid.
Its basic properties in position space are
similar to the ones known from the spatially local theory,
in particular, since
$\tensor{Q}(\mathbf{r},\mathbf{r}',\omega)$ is reciprocal, so is
$\tensor{G}(\mathbf{r},\mathbf{r}',\omega)$,
$\tensor{G}(\mathbf{r},\mathbf{r}',\omega)\!=\!
\tensor{G}{^\mathsf{T}}(\mathbf{r}',\mathbf{r},\omega)$,
and, for real $\omega$, the generalized integral relation
\begin{multline}
\label{eq8-4}
\mu_{0}\omega
\int \D^3s\int \D^3s' \,\tensor{G}(\mathbf{r,s},\omega)
\cdot
\tensor{\sigma}(\mathbf{s,s'},\omega)
\cdot\tensor{G}{^{\ast}}(\mathbf{s',r'},\omega)
\\
= \Im\tensor{G}(\mathbf{r,r'},\omega)
\end{multline}
holds (App.~\ref{AppA}).

To quantize the theory,
the Langevin noise source $\fon{j}_{\mathrm N}(\mathbf{r},\omega)$
is regarded as an operator
[$\fon{j}_{\mathrm N}(\mathbf{r},\omega)\mapsto
\fo{j}_{\mathrm N}(\mathbf{r},\omega)$]
with the commutation relation
\begin{equation}
\label{eq8-6}
\bigl[\fo{j}_{\mathrm N}(\mathbf{r},\omega),
\fo{j}{^{\dagger}_{\mathrm N}}(\mathbf{r'},\omega')\bigr]
=
\frac{\hbar\omega}{\pi}
\,\delta(\omega-\omega')\,
\tensor{\sigma}(\mathbf{r},\mathbf{r}',\omega).
\end{equation}
The (now operator-valued) equation (\ref{NL7})
relates the electric field operator
\begin{equation}
\label{eq13}
\hat{\mathbf{E}}(\mathbf{r})
=\int_{0}^{\infty}\D\omega\,\fo{E}(\mathbf{r},\omega) + \mathrm{H.c.},
\end{equation}
and thus all the electromagnetic field operators,
to
$\fo{j}_{\mathrm N}(\mathbf{r},\omega)$ and
$\fo{j}{^{\dagger}_{\mathrm N}}(\mathbf{r'},\omega')$,
which may be regarded as the dynamical variables of the overall
system consisting of the electromagnetic field and the linear medium
(incorporating the reservoir degrees of freedom
responsible for absorption).
It should be mentioned that,
by means of the correspondence
\begin{equation}
\label{eq8-6-1}
\frac{i}{\varepsilon_{0}\omega}\,
\tensor{Q}(\mathbf{r},\mathbf{r}',\omega)\leftrightarrow
\tensor{\chi}(\mathbf{r},\mathbf{r}',\omega),
\end{equation}
where $\tensor{\chi}(\mathbf{r},\mathbf{r}',\omega)$ is
the (nonlocal) dielectric susceptibility tensor,
the basic commutation relation (\ref{eq8-6})
becomes equivalent to the commutation relation
derived from a microscopic, linear two-band model of
dielectric material \cite{DiStefanoO082001},
which has been used to study the quantized
electromagnetic field in spatially dispersive dielectrics
\cite{SavastaS032002,SavastaS082002}.

In order to complete the quantization scheme,
a Hamiltonian $\hat{H}$
needs to be introduced [as a functional of
$\fo{j}_{\mathrm N}(\mathbf{r},\omega)$ and
$\fo{j}{^{\dagger}_{\mathrm N}}(\mathbf{r'},\omega')$] so as to
generate `free' time evolution
according to
\begin{equation}
\label{NL30}
[\fo{j}_{\mathrm N}(\mathbf{r},\omega),\hat{H}]=\hbar\omega\,
\fo{j}_{\mathrm N}(\mathbf{r},\omega),
\end{equation}
which constrains the Hamiltonian to the form
\begin{equation}
\label{NL28-0}
\hat H
=
\pi\!
\int_{0}^{\infty}
\D\omega\!
\int \D^3r\!
\int \D^3r'\,
\fo{j}_{\mathrm N}^{\dagger}(\mathbf{r},\omega)
\cdot
\tensor{\rho}(\mathbf{r,r'},\omega)
\cdot
\fo{j}_{\mathrm N}(\mathbf{r'},\omega),
\end{equation}
to within irrelevant c-number contributions.
A glance at Eqs.~(\ref{eq8-6}) and (\ref{NL30})
now shows that
$\tensor{\rho}(\mathbf{r,r'},\omega)$ is
the integral kernel of the inverse operator
of the operator associated with $\tensor{\sigma}(\mathbf{r,r'},\omega)$
(which exists).
The validity of the quantization scheme is confirmed by
checking that the well-known
equal-time commutation relations for the electromagnetic
field operators hold, which can be done, in analogy to
the spatially local theory
(cf.~Refs.~\cite{KnoellL002001,ScheelS071998,HoTD102003}),
by properly taking into account
the properties of
$\tensor{G}(\mathbf{r},\mathbf{r}',\omega)$ and
$\tensor{Q}(\mathbf{r},\mathbf{r}',\omega)$
[in particular, Eq.~(\ref{eq8-4})].

The Hamiltonian (\ref{NL28-0}) may clearly be brought to
the diagonal form
\begin{equation}
\label{NL28}
\hat H
=\int \D^3r \int_{0}^{\infty}
\D\omega\, \hbar\omega\,
\hat{\mathbf{f}}^{\dagger}(\mathbf{r},\omega)
\cdot
\hat{\mathbf{f}}(\mathbf{r},\omega)
\end{equation}
known from the spatially local theory,
where $\hat{\mathbf{f}}(\mathbf{r},\omega)$ is a bosonic field,
\begin{equation}
\label{eq8-7a}
\bigl[\hat{\mathbf{f}}(\mathbf{r},\omega),\hat{\mathbf{f}}^{\dagger}
(\mathbf{r'},\omega')\bigr]
=
\delta(\omega-\omega')\tensor{I}\delta(\mathbf{r}-\mathbf{r}'),
\end{equation}
by performing a linear transformation of the variables,
which we shall assume to be invertible.
Writing
\begin{equation}
\label{eq8-5}
\fo{j}_{\mathrm N}(\mathbf{r},\omega)
=
\left(\frac{\hbar\omega}{\pi}\right)^{\frac{1}{2}}
\int \D^3r'\, \tensor{K}(\mathbf{r},\mathbf{r}',\omega)
\cdot
\hat{\mathbf{f}}(\mathbf{r}',\omega),
\end{equation}
the diagonalization is achieved and
Eqs.~(\ref{eq8-7a}) and (\ref{eq8-6}) are rendered equivalent
if we choose the integral kernel
$\tensor{K}(\mathbf{r},\mathbf{r}',\omega)$
such that, for real $\omega$, the integral equation
\begin{equation}
\label{eq8-7}
\int \D^3s\,
\tensor{K}(\mathbf{r,s},\omega) \cdot
\tensor{K}{^{+}}(\mathbf{r',s},\omega)
=
\tensor{\sigma}(\mathbf{r,r'},\omega)
\end{equation}
holds, which is guaranteed to possess
solutions (see Sec.~\ref{sec3})
since $\tensor{\sigma}(\mathbf{r},\mathbf{r}',\omega)$ is the
integral kernel of a positive definite operator.

So far we have considered the `free' medium-assisted
electromagnetic field. Its interaction with additional
(e.g., atomic) systems
can be included in the theory on the basis of
the well-known minimal or multi-polar coupling schemes in
the usual way (see, e.g., Ref.~\cite{WelschQuantumOptics}).

%
\section{Natural variables and projective variables}
\label{sec3}

Let us now turn to the problem of constructing
the integral kernel $\tensor{K}(\mathbf{r},\mathbf{r'},\omega)$
in Eq.~(\ref{eq8-7}). For this purpose, we consider the eigenvalue problem
\begin{equation}
\label{NL16}
\int \D^3r'\,
\tensor{\sigma}(\mathbf{r},\mathbf{r}',\omega)
\cdot
\mathbf{F}(\alpha,\mathbf{r}',\omega)
=
\sigma(\alpha,\omega)
\mathbf{F}(\alpha,\mathbf{r},\omega)
\end{equation}
which, under appropriate regularity assumptions on the conductivity
tensor $\tensor{Q}(\mathbf{r},\mathbf{r}',\omega)$,
such as those listed below Eqs.~(\ref{eq8-1}), is well-defined.
In particular, it features a real (positive) spectrum
and a complete set of orthogonal eigensolutions,
which we may take to be ($\delta$-)normalized. Note that
the real $\omega$ plays the role of a parameter here, and $\alpha$
stands for the collection of (discrete and/or continuous) indices
needed to label the eigenfunctions. Adopting a continu\-um 
notation, we may write
\begin{align}
\label{NL18}
&
\int \D\alpha\,
\mathbf{F}(\alpha,\mathbf{r},\omega)
\ot
\mathbf{F}^{\ast}(\alpha,\mathbf{r}',\omega)
=
\tensor{I}\delta(\mathbf{r}-\mathbf{r}'),
\\[1ex]
\label{NL19}
&
\int \D^3r\,\mathbf{F}^{\ast}
(\alpha,\mathbf{r},\omega)\cdot
\mathbf{F}(\alpha',\mathbf{r},\omega)
=
\delta(\alpha-\alpha'),
\end{align}
and the diagonal expansion of
$\tensor{\sigma}(\mathbf{r},\mathbf{r}',\omega)$ reads
\begin{equation}
\label{NL20}
\tensor{\sigma}(\mathbf{r},\mathbf{r}',\omega)
=\int \D\alpha\, \sigma(\alpha,\omega)
\mathbf{F}(\alpha,\mathbf{r},\omega)
\ot
\mathbf{F}^{\ast}(\alpha,\mathbf{r}',\omega),
\end{equation}
which resembles the expansion of the dielectric susceptibility
in the microscopic theory \cite{DiStefanoO082001} mentioned above.
Substituting Eq.~(\ref{NL20}) into Eq.~(\ref{eq8-7}),
we may construct an integral kernel
$\tensor{K}(\mathbf{r},\mathbf{r}',\omega)$ in the form of
\begin{equation}
\label{NL21}
\tensor{K}(\mathbf{r},\mathbf{r}',\omega)
=
\int \D\alpha\,
\sigma^{\frac{1}{2}}(\alpha,\omega)\,
\mathbf{F}(\alpha,\mathbf{r},\omega)
\ot
\mathbf{F}^{\ast}(\alpha,\mathbf{r}',\omega),
\end{equation}
where we choose
$\sigma^{1/2}(\alpha,\omega)$ $\!>$ $\!0$
so that the operator associated with
$\tensor{K}(\mathbf{r},\mathbf{r}',\omega)$ is
the positive, Hermitian square-root of the operator associated with
$\tensor{\sigma}(\mathbf{r},\mathbf{r}',\omega)$.
Obviously, this solution to Eq.~(\ref{eq8-7}) is not
unique, since any other kernel of the form
\begin{equation}
\label{NL21-1}
\tensor{K}{'}(\mathbf{r},\mathbf{r}',\omega)
=
\int \D^3s\,\tensor{K}(\mathbf{r,s},\omega)
\cdot
\tensor{V}(\mathbf{s,r'},\omega)
\end{equation}
with $\tensor{V}(\mathbf{r},\mathbf{s},\omega)$
satisfying
\begin{equation}
\label{NL21-2}
\int \D^3s\,
\tensor{V}(\mathbf{r},\mathbf{s},\omega) \cdot
\tensor{V}{^{+}}(\mathbf{r}',\mathbf{s},\omega)
=\tensor{I}\delta(\mathbf{r}-\mathbf{r}')
\end{equation}
also obeys Eq.~(\ref{eq8-7}).
As we are interested in invertible transformations (\ref{eq8-5}),
the operator corresponding to
$\tensor{V}(\mathbf{r},\mathbf{s},\omega)$ should be invertible as well,
so that we can replace Eq.~(\ref{NL21-2}) with the
stronger unitarity condition
\begin{multline}
\label{B3}
\int \D^3s\,
\tensor{V}{^+}(\mathbf{s},\mathbf{r},\omega)
\cdot
\tensor{V}(\mathbf{s},\mathbf{r}',\omega)
\\
= \int \D^3s\,
\tensor{V}(\mathbf{r},\mathbf{s},\omega)
\cdot
\tensor{V}{^+}(\mathbf{r}',\mathbf{s},\omega)
=\tensor{I}\delta(\mathbf{r}-\mathbf{r}').
\end{multline}
Without loss of generality
(see App.~\ref{AppB}), we can base our
further calculations on Eq.~(\ref{NL21}).

Inserting Eq.~(\ref{NL21}) into Eq.~(\ref{eq8-5}), we find that
\begin{equation}
\label{NL23}
\fo{j}_{\mathrm N}(\mathbf{r},\omega)
=
\left(\frac{\hbar\omega}{\pi}\right)^{\frac{1}{2}}
\!\!
\int \D\alpha\,
\sigma^{\frac{1}{2}}(\alpha,\omega)\,
\mathbf{F}(\alpha,\mathbf{r},\omega)
\hat{g}(\alpha,\omega),
\end{equation}
where we have introduced the new variables
\begin{equation}
\label{NL25}
\hat{{g}}(\alpha,\omega)
=
\int \D^3r\,
\mathbf{F}^{\ast}(\alpha,\mathbf{r},\omega)
\cdot
\hat{\mathbf{f}}(\mathbf{r},\omega) \,,
\end{equation}
referred to as the natural variables in the following.
Needless to say that they are again of bosonic type,
\begin{align}
\label{NL27}
\bigl[
\hat{{g}}(\alpha,\omega),
\hat{{g}}^{\dagger}(\alpha',\omega')
\bigr]
=
\delta(\alpha-\alpha')\delta(\omega-\omega').
\end{align}
Since the transformation (\ref{NL25}) does
not mix different $\omega$~components,
the Hamiltonian (\ref{NL28}) is still diagonal when expressed
in terms of the natural variables,
\begin{equation}
\label{NL29}
\hat H
=
\int\D\alpha \int_{0}^{\infty}
\D\omega\, \hbar\omega \,
\hat{g}^{\dagger}(\alpha,\omega)
\hat{{g}}(\alpha,\omega),
\end{equation}
as can be easily seen by inverting Eq.~(\ref{NL25}),
\begin{equation}
\label{NL25a}
\hat{\mathbf{f}}(\mathbf{r},\omega)
=
\int \D\alpha\,
\mathbf{F}(\alpha,\mathbf{r},\omega)
\hat{{g}}(\alpha,\omega),
\end{equation}
and combining with Eq.~(\ref{NL28}), on recalling Eq.~(\ref{NL19}).

Let us organize the set of eigenfunctions
$\mathbf{F}(\alpha,\mathbf{r},\omega)$
into (a discrete number of)
subsets labeled by $\lambda$ ($\lambda$ $\!=$ $\!1,2,\ldots,\Lambda$).
With the notation \mbox{$\alpha \mapsto (\lambda,\beta)$},
Eq.~(\ref{NL25a}) then reads
\begin{equation}
\label{NL25a-1}
\hat{\mathbf{f}}(\mathbf{r},\omega)
=
\sum_{\lambda}\hat{\mathbf{f}}_{\lambda}(\mathbf{r},\omega),
\end{equation}
where
\begin{equation}
\label{NL25a-2}
\hat{\mathbf{f}}_{\lambda}(\mathbf{r},\omega)
=
\int \D\beta\,
\mathbf{F}_{\lambda}(\beta,\mathbf{r},\omega)
\hat{{g}}_{\lambda}(\beta,\omega).
\end{equation}
The operators associated with the integral kernels
\begin{equation}
\label{NL18-1}
\tensor{P}_{\lambda}(\mathbf{r},\mathbf{r}',\omega)
=
\int \D\beta\,
\mathbf{F}_{\lambda}(\beta,\mathbf{r},\omega)
\ot
\mathbf{F}_{\lambda}^{\ast}(\beta,\mathbf{r}',\omega)
\end{equation}
form a complete set of orthogonal projectors.
Obviously, these projectors and the operators associated with
$\tensor{\sigma}(\mathbf{r},\mathbf{r}',\omega)$
and $\tensor{K}(\mathbf{r},\mathbf{r}',\omega)$ as given by
Eq.~(\ref{NL21}) are commuting quantities.
It is not difficult to see that the variables
\begin{align}
\label{NL25d}
\hat{\mathbf{f}}_{\lambda}(\mathbf{r},\omega)
&
=
\int\D^3r'\,
\tensor{P}_{\lambda}(\mathbf{r},\mathbf{r}',\omega)
\cdot
\hat{\mathbf{f}}(\mathbf{r}',\omega)
\nonumber \\
&
=
\int \D\beta\,
\mathbf{F}_{\lambda}(\beta,\mathbf{r},\omega)
\hat{{g}}_{\lambda}(\beta,\omega),
\end{align}
referred to as projective variables in the following,
obey the non-bosonic commutation relation
\begin{equation}
\label{eq8-16}
\bigl[
\hat{\mathbf{f}}_{\lambda}(\mathbf{r},\omega),
\hat{\mathbf{f}}_{\lambda'}^{\dagger}(\mathbf{r'},\omega')
\bigr]
=
\delta_{\lambda\lambda'}\delta(\omega-\omega')
\tensor{P}_{\lambda}(\mathbf{r,r'},\omega),
\end{equation}
and the Hamiltonian (\ref{NL28}) expressed in terms of the projective
variables reads as
\begin{equation}
\label{NL28-1}
\hat H
=
\sum_{\lambda} \int \D^3r
\int_{0}^{\infty}
\D\omega\, \hbar\omega\,
\hat{\mathbf{f}}^{\dagger}_{\lambda}(\mathbf{r},\omega)
\cdot
\hat{\mathbf{f}}_{\lambda}(\mathbf{r},\omega).
\end{equation}
From Eqs.(\ref{eq8-16}) and (\ref{NL28-1}) it then follows that
\begin{align}
\label{NL28-2}
\bigl[
\hat{\mathbf{f}}_{\lambda}(\mathbf{r},\omega),\hat H
\bigr]
&
=
\hbar\omega \!\!\int\! \D^3r'\,
\tensor{P}_{\lambda}(\mathbf{r},\mathbf{r}',\omega)
\cdot
\hat{\mathbf{f}}_{\lambda}(\mathbf{r}',\omega)
\nonumber\\
&
=
\hbar\omega\,
\hat{\mathbf{f}}_{\lambda}(\mathbf{r},\omega).
\end{align}
Inserting Eq.~(\ref{NL25a-1}) in Eq.~(\ref{eq8-5}), we obtain
\begin{equation}
\label{eq8-5-1}
\fo{j}_{\mathrm N}(\mathbf{r},\omega)
=
\sum_{\lambda}\,
\fo{j}_{\mathrm N\lambda}(\mathbf{r},\omega),
\end{equation}
where the $\fo{j}_{\mathrm N\lambda}(\mathbf{r},\omega)$
are given by
\begin{equation}
\label{eq8-5c}
\fo{j}_{\mathrm N\lambda}(\mathbf{r},\omega)
=
\left(\frac{\hbar\omega}{\pi}\right)^{\frac{1}{2}}
\int \D^3r'\,\tensor{K}_{\lambda}
(\mathbf{r},\mathbf{r}',\omega)
\cdot
\hat{\mathbf{f}}_{\lambda}(\mathbf{r}',\omega),
\end{equation}
with
\begin{align}
\label{eq9-1}
\tensor{K}_{\lambda}(\mathbf{r},\mathbf{r}',\omega)
&
=
\int\D^3s\,
\tensor{P}_{\lambda}(\mathbf{r,s},\omega)
\cdot
\tensor{K}(\mathbf{s,r'},\omega)
\nonumber\\
&
=
\int\D^3s\,
\tensor{K}(\mathbf{r},\mathbf{s},\omega)
\cdot
\tensor{P}_{\lambda}(\mathbf{s},\mathbf{r}',\omega).
\end{align}
Recalling Eq.~(\ref{eq8-16}), we can easily see that
\begin{equation}
\label{eq8-6a}
\bigl[\fo{j}_{\mathrm N\lambda}(\mathbf{r},\omega),
\fo{j}{^{\dagger}_{\mathrm N\lambda'}}(\mathbf{r'},\omega')\bigr]
=\frac{\hbar\omega}{\pi}
\delta_{\lambda\lambda'}\delta(\omega-\omega')
\tensor{\sigma}_{\lambda}
(\mathbf{r,r'},\omega),
\end{equation}
where $\tensor{\sigma}_{\lambda}(\mathbf{r},\mathbf{r}',\omega)$
is defined according to Eq.~(\ref{eq9-1}) with
$\tensor{\sigma}$ in place of $\tensor{K}$. Summation of
Eq.~(\ref{eq8-6a}) over $\lambda$ and $\lambda'$ leads back to
Eq.~(\ref{eq8-6}),
so that the two equations are equivalent.

At this stage, we observe that
there is the option
to base the quantization scheme directly on
Eqs.~(\ref{NL28-1}), (\ref{eq8-5-1}), and (\ref{eq8-5c}),
regarding the variables
$\hat{\mathbf{f}}_{\lambda}(\mathbf{r},\omega)$ and
$\hat{\mathbf{f}}_{\lambda}^{\dagger}(\mathbf{r},\omega)$
as the basic dynamical variables of the theory and
assigning to them
bosonic commutation relations
\begin{equation}
\label{eq8-16-1}
\bigl[
\hat{\mathbf{f}}_{\lambda}
(\mathbf{r},\omega),
\hat{\mathbf{f}}_{\lambda'}^{\dagger}(\mathbf{r'},\omega')
\bigr]
=
\delta_{\lambda\lambda'}\delta(\omega-\omega')
\tensor{I}\delta(\mathbf{r-r'})
\end{equation}
in place of Eq.~(\ref{eq8-16}). Note that, in so doing,
back reference from the variables
$\hat{\mathbf{f}}_\lambda(\mathbf{r},\omega)$ to 
the original variables $\hat{\mathbf{f}}(\mathbf{r},\omega)$
is not possible anymore.
As can be seen from Eqs.~(\ref{eq8-5c}) and (\ref{eq9-1}),
Eq.~(\ref{eq8-6a}) is satisfied also
when the $\hat{\mathbf{f}}_{\lambda}(\mathbf{r},\omega)$
and $\hat{\mathbf{f}}_{\lambda}^{\dagger}(\mathbf{r},\omega)$
are considered as bosonic variables, from which it follows
[via Eq.~(\ref{eq8-5-1})] that Eq.~(\ref{eq8-6}) also
still holds and, as before, this implies that
the correct electromagnetic-field
commutation relations hold. The second line of Eq.~(\ref{NL28-2})
remains of course also true so that the correct time evolution
is ensured as well.
%

%
Since the state space attributed to the bosonic variables
$\hat{\mathbf{f}}_\lambda(\mathbf{r},\omega)$ and
$\hat{\mathbf{f}}_\lambda^\dagger(\mathbf{r},\omega)$
is, in general, different from the
state space attributed to the original variables
$\hat{\mathbf{f}}(\mathbf{r},\omega)$ and
$\hat{\mathbf{f}}^\dagger(\mathbf{r},\omega)$
[or, equivalently, attributed to $\hat{g}_{\lambda}(\beta,\omega)$
and $\hat{g}_{\lambda}^\dagger(\beta,\omega)$],
the allowable states must be restricted, by ruling out
certain coherent superpositions of states in
the sense of a super-selection rule.
In App.~\ref{AppC}, we show that the
condition imposed on the states
may be described by means of
a set of projectors $\hat{P}_{\lambda}$ such
that the allowable states $\ket{\psi}$ can be
characterized by
\begin{equation}
\label{eq8-16-2}
\hat{P}_{\lambda}\ket{\psi}
=
\ket{\psi} \ \forall\, \lambda,
\end{equation}
where the action of the projectors
$\hat{P}_{\lambda}$ in state space
is closely related to
the action of the projectors associated with the
kernels (\ref{NL18-1}) in position space.
As a result, if the total
Hamiltonian $\hat{H}_\mathrm{tot}$
composed of the Hamiltonian (\ref{NL28-1})
and possible interaction terms
(in the case where additional, active sources are present)
commutes with all of the projectors $\hat{P}_{\lambda}$,
\begin{equation}
\label{eq8-16-3}
\bigl[\hat{P}_{\lambda},\hat{H}_\mathrm{tot}\bigr]=0
 \ \forall\, \lambda
,
\end{equation}
then allowable states remain allowable in the course of time, and
the option of treating the
$\hat{\mathbf{f}}_\lambda(\mathbf{r},\omega)$ and
$\hat{\mathbf{f}}_\lambda^\dagger(\mathbf{r},\omega)$
as bosonic variables can be safely exercised.
Clearly, all the observables of interest should then
also commute with the $\hat{P}_{\lambda}$ so that
no transition matrix elements
between states belonging to different subspaces, i.e.,
between spaces attributed
to different $\lambda$ values,
can ever come into play.

One can also consider decompositions
of $\fo{j}_{\mathrm N}(\mathbf{r},\omega)$, where
in place of the
$\fo{j}_{\mathrm N\lambda}(\mathbf{r},\omega)$
introduced above other
quantities
$\fo{J}_{\mathrm N\lambda}(\mathbf{r},\omega)$
subject to the condition
\begin{equation}
\label{eq8-101}
\sum_{\lambda} \fo{J}_{\mathrm N\lambda}(\mathbf{r},\omega)
=
\sum_{\lambda} \fo{j}_{\mathrm N\lambda}(\mathbf{r},\omega)
\end{equation}
are introduced, whose commutation relations may be quite different
from those of the $\fo{j}_{\mathrm N\lambda}(\mathbf{r},\omega)$.
Obviously, the total noise current density
$\fo{j}_{\mathrm N}(\mathbf{r},\omega)$ as defined by
Eq.~(\ref{eq8-5-1}) and the commutation relation (\ref{eq8-6})
are not changed by such a transformation,
briefly
referred to as gauge transformation in the following.
Moreover, since, with regard to Eq.~(\ref{eq8-6}),
only the sum of the commutators
$[\fo{J}_{\mathrm N\lambda}(\mathbf{r},\omega),
\fo{J}{_{\mathrm N\lambda'}^{\dagger}}(\mathbf{r}',\omega')]$
over all $\lambda$ and $\lambda'$ is relevant, 
every chosen set of (algebraically consistent) commutators
$[\fo{J}_{\mathrm N\lambda}(\mathbf{r},\omega),
\fo{J}{_{\mathrm N\lambda'}^{\dagger}}(\mathbf{r}',\omega')]$
which leads to Eq.~(\ref{eq8-6})
yields, in principle, a consistent
quantization scheme in its own right.
A `substructure below' Eq.~(\ref{eq8-6}) can hence be
introduced with some arbitrariness, but since the various
available alternatives are not necessarily equivalent
to each other, a specific one should not be favored in the absence of
good (physical) motivation.
In contrast, if the observables of interest---including
the Hamiltonian---can be viewed as functionals of
$\fo{j}_\mathrm N(\mathbf{r},\omega)$
[rather than of the individual
$\fo{J}_{\mathrm N\lambda}(\mathbf{r},\omega)$],
Eqs.~(\ref{eq8-6}) and (\ref{NL28-0})
can be regarded,
in view of the fluctuation-dissipation
theorem(s) (see, e.g., Ref.~\cite{KuboNonEqStat}),
as being unique, and hence, as invariable fundament of the theory.

From the above, it may be reasonable to widen the notion 
of projective variables as follows.
If, for a chosen (physically motivated)
decomposition of the noise current density,
it is possible to (linearly) relate the
$\fo{J}_{\mathrm N\lambda}(\mathbf{r},\omega)$
in Eq.~(\ref{eq8-101})
to (new) variables
$\hat{\mathbf{f}}_\lambda(\mathbf{r},\omega)$
such that, upon
considering the latter as bosonic variables,
the validity of the basic equations
(\ref{eq8-6}) and (\ref{NL30}) is ensured,
then the specific
quantization scheme so obtained
may be regarded as arising
from the general quantization scheme
by excluding certain types of (superposition) states from state space, and
restricting the dynamics (as well as
the allowable observables) accordingly.
The $\hat{\mathbf{f}}_\lambda(\mathbf{r},\omega)$
may then be seen as projective variables in a wider sense.
%


\section{Different classes of media}
\label{sec4}

We proceed to show that rather different classes of media
(usually studied separately) fit into
the general quantization scheme
developed in Sec.~\ref{sec2}.
The main task to be performed is solving the
eigenvalue problem (\ref{NL16}), which requires
knowledge of $\tensor{\sigma}(\mathbf{r},\mathbf{r}',\omega)$
for the specific medium under consideration.
In two limiting cases, the
exact solution to Eq.~(\ref{NL16}) can be given straightforwardly,
namely, in the case of an inhomogeneous
medium without spatial dispersion
and in the case of a
homogeneous medium that shows spatial dispersion.
Let us, therefore, first examine
these two cases in detail before considering
more general situations.


\subsection{Spatially non-dispersive inhomogeneous media}
\label{sec4B}

The complete neglect of spatial dispersion
means to regard the medium response, i.e.,
$\tensor{Q}(\mathbf{r},\mathbf{r}',\omega)$,
as being strictly local. If this is assumed, we have
\begin{equation}
\label{NLB1}
\tensor{\sigma}(\mathbf{r},\mathbf{r}',\omega)
=
\tensor{\sigma}(\mathbf{r},\omega)
\delta(\mathbf{r}-\mathbf{r}'),
\end{equation}
where $\tensor{\sigma}(\mathbf{r},\omega)$
can be written in diagonal form as
\begin{equation}
\label{NLB1a}
\tensor{\sigma}(\mathbf{r},\omega)
=
\sum_{i=1}^{3}
\sigma_{i}(\mathbf{r},\omega)\,
\mathbf{e}_{i}(\mathbf{r},\omega)
\ot
\mathbf{e}_{i}^\ast(\mathbf{r},\omega),
\end{equation}
with $\mathbf{e}_{i}(\mathbf{r},\omega)$
($i$ $\!=$ $\!1,2,3$) being orthonormal unit vectors.
Hence, the eigenvalues $\sigma(\alpha,\omega)$ and
eigenfunctions $\mathbf{F}(\alpha,\mathbf{r},\omega)$ of the operator
associated with $\tensor{\sigma}(\mathbf{r},\mathbf{r}',\omega)$
read [$\alpha\mapsto(i,\mathbf{s})$]
$\sigma_{i}(\mathbf{s},\omega)$ and
\begin{equation}
\label{NLB3}
\mathbf{F}_{i} (\mathbf{s},\mathbf{r},\omega)
=
\mathbf{e}_{i}(\mathbf{s},\omega)
\delta(\mathbf{s}-\mathbf{r}),
\end{equation}
respectively. Equation~(\ref{NL21}) then becomes
\begin{equation}
\label{NLB6}
\tensor{K}(\mathbf{r},\mathbf{r}',\omega)
=
\tensor{K}(\mathbf{r},\omega)
\delta(\mathbf{r}-\mathbf{r}'),
\end{equation}
where
\begin{equation}
\label{NLB6-1}
\tensor{K}(\mathbf{r},\omega)
=
\sum_{i=1}^{3}
\sigma_{i}^{1/2}(\mathbf{r},\omega)\,
\mathbf{e}_{i}(\mathbf{r},\omega)
\ot
\mathbf{e}_{i}^{\ast}(\mathbf{r},\omega),
\end{equation}
and Eq.~(\ref{eq8-5}) takes the form
\begin{equation}
\label{NLB8-3}
\fo{j}_{\mathrm N}(\mathbf{r},\omega)
=
\left(\frac{\hbar\omega}{\pi}\right)^{\frac{1}{2}}
\tensor{K}(\mathbf{r},\omega)
\cdot
\hat{\mathbf{f}}(\mathbf{r},\omega),
\end{equation}
which just yields the well-known quantization scheme
for a locally responding,
possibly anisotropic dielectric material
\cite{ScheelS071998,KnoellL002001},
upon identifying
$\tensor{\sigma}(\mathbf{r},\omega)$ $\!=$
$\!\varepsilon_{0}\omega\,\Im\tensor{\chi}(\mathbf{r},\omega)$,
with $\tensor{\chi}(\mathbf{r},\omega)$ being the (local)
dielectric susceptibility tensor [cf.~Eq.~(\ref{eq8-6-1})].
The natural variables $\hat{{g}}_{i}(\mathbf{r},\omega)$ are
here simply the components of $\hat{\mathbf{f}}(\mathbf{r},\omega)$
along the principal axes of the medium,
which may in general vary with position and frequency,
\begin{align}
\label{NLB7}
&
\hat{{g}}_{i}(\mathbf{r},\omega)
=
\mathbf{e}^{\ast}_{i}(\mathbf{r},\omega)
\cdot
\hat{\mathbf{f}}(\mathbf{r},\omega),
\\[1ex]
\label{NLB8}
&
\hat{\mathbf{f}}(\mathbf{r},\omega)
=\sum_{i=1}^{3}
\mathbf{e}_{i}(\mathbf{r},\omega)
\hat{{g}}_{i}(\mathbf{r},\omega).
\end{align}
Identifying the index $\lambda$ introduced in Eq.~(\ref{NL25a-1})
with $i$ and assuming that
$\sigma_{i}(\mathbf{r},\omega)$ $\!\neq$
$\!\sigma_{i'}(\mathbf{r},\omega)$ for
$i$ $\!\neq$ $\!i'$, one can define,
according to Eq.~(\ref{NL18-1}),
the three projection kernels
\begin{equation}
\label{NLB8-1}
\tensor{P}_{i}(\mathbf{r},\mathbf{r}',\omega)
=
\mathbf{e}_{i}(\mathbf{r},\omega)
\ot
\mathbf{e}_{i}^{\ast}(\mathbf{r}',\omega)
\delta(\mathbf{r}-\mathbf{r}'),
\end{equation}
which, according to Eq.~(\ref{NL25d}), give
rise to three sets of projective
variables,
\begin{equation}
\label{NLB8-2}
\hat{\mathbf{f}}_{i}(\mathbf{r},\omega)
=
\mathbf{e}_{i}(\mathbf{r},\omega)
\hat{{g}}_{i}(\mathbf{r},\omega).
\end{equation}
As long as the projective variables are not coupled to each
other---which is obviously the case for the
`free' system governed by the Hamiltonian (\ref{NL28-2})---they can be
regarded as being of bosonic type. In this case, instead of using the
original set of bosonic variables
$\hat{\mathbf{f}}(\mathbf{r},\omega)$ and
$\hat{\mathbf{f}}^\dagger(\mathbf{r},\omega)$,
one can use three sets of bosonic variables
$\hat{\mathbf{f}}_{i}(\mathbf{r},\omega)$ and
$\hat{\mathbf{f}}_{i}^\dagger(\mathbf{r},\omega)$
associated with the three
principal axes of the dielectric medium
at each space point.

If two of the three eigenvalues
$\sigma_{i}(\mathbf{r},\omega)$
coincide (uniaxial medium),
the two corresponding
projection kernels $\tensor{P}_{i}(\mathbf{r},\mathbf{r}',\omega)$
should be combined into one projector
(projecting on the plane perpendicular to
the distinguished axis of the medium),
thereby reducing the number of sets of projective variables to two.
Clearly, if the three eigenvalues
$\sigma_{i}(\mathbf{r},\omega)$ all
coincide (isotropic medium), the three projection kernels
$\tensor{P}_{i}(\mathbf{r},\mathbf{r}',\omega)$
should be combined to give the
unit kernel $\tensor{I}\delta(\mathbf{r}-\mathbf{r}')$,
corresponding to the use of the original variables.

%
\subsection{Spatially dispersive homogeneous media}
\label{sec4C}

In the limiting case of an (infinitely extended) homogeneous medium,
$\tensor{Q}(\mathbf{r},\mathbf{r}',\omega)$ is
translationally invariant, i.e., it is a function of the difference
$\mathbf{r}\!-\!\mathbf{r}'$, and so is then
$\tensor{\sigma}(\mathbf{r},\mathbf{r}',\omega)$.
We may therefore represent it as the spatial Fourier transform
\begin{equation}
\label{eq8-20}
\tensor{\sigma}(\mathbf{r},\mathbf{r}',\omega)
=
\frac{1}{(2\pi)^{3}}
\int \D^3k\,\tensor{\sigma}(\mathbf{k},\omega)
e^{i\mathbf{k}\cdot(\mathbf{r}-\mathbf{r}')},
\end{equation}
where
\begin{equation}
\label{eq8-21}
\tensor{\sigma}(\mathbf{k},\omega)
=
\sum_{i=1}^{3}
\sigma_{i}(\mathbf{k},\omega)\,
\mathbf{e}_{i}(\mathbf{k},\omega)
\ot
\mathbf{e}_{i}^{\ast}(\mathbf{k},\omega),
\end{equation}
with $\mathbf{e}_{i}(\mathbf{k},\omega)$
($i$ $\!=$ $\!1,2,3$) being orthogonal unit vectors.
Consequently, the eigenvalues
$\sigma(\alpha,\omega)$ and eigenfunctions
$\mathbf{F}(\alpha,\mathbf{r},\omega)$
of the operator
associated with $\tensor{\sigma}(\mathbf{r},\mathbf{r}',\omega)$
are [$\alpha\mapsto(i,\mathbf{k})$]
$\sigma_{i}(\mathbf{k},\omega)$ and
\begin{equation}
\label{eq8-22}
\mathbf{F}_{i}(\mathbf{k},\mathbf{r},\omega)
=
(2\pi)^{-3/2}
e^{i\mathbf{k}\cdot\mathbf{r}}
\mathbf{e}_{i}(\mathbf{k},\omega),
\end{equation}
respectively, and Eq.~(\ref{NL21}) reads
\begin{equation}
\label{NLB12}
\tensor{K}(\mathbf{r},\mathbf{r}',\omega)
=
\frac{1}{(2\pi)^{3}}
\int \D^3k\,
\tensor{K}(\mathbf{k},\omega)
e^{i\mathbf{k}\cdot(\mathbf{r}-\mathbf{r}')},
\end{equation}
where
\begin{equation}
\label{eq120}
\tensor{K}(\mathbf{k},\omega)
=
\sum_{i=1}^{3}
\sigma_{i}^{1/2}(\mathbf{k},\omega)\,
\mathbf{e}_{i}(\mathbf{k},\omega)
\ot
\mathbf{e}_{i}^{\ast}(\mathbf{k},\omega).
\end{equation}
Combination of Eqs.~(\ref{eq8-5}), (\ref{NLB12}), and (\ref{eq120})
then yields
\begin{multline}
\label{eq120-1}
\fon{\mathbf{j}}_\mathrm{N}(\mathbf{r},\omega)
=
\left(\frac{\hbar\omega}{\pi}\right)^{\frac{1}{2}}
\frac{1}{(2\pi)^{3/2}}
\\\times\,
\sum_{i=1}^3\int\D^3k\,
e^{i\mathbf{k}\cdot\mathbf{r}}
\sigma_{i}^{1/2}(\mathbf{k},\omega)
\mathbf{e}_{i}(\mathbf{k},\omega)
\hat{g}_{i}(\mathbf{k},\omega),
\end{multline}
where the natural variables
$\hat{g}_{i}(\mathbf{k},\omega)$
are related to the spatial Fourier components of
$\hat{\mathbf{f}}(\mathbf{r},\omega)$ as
\begin{equation}
\label{NLB13}
\hat{{g}}_{i}(\mathbf{k},\omega)
=
\frac{1}{(2\pi)^{3/2}}
\int \D^3r\,
e^{-i\mathbf{k}\cdot \mathbf{r}}\,
\mathbf{e}_{i}^{\ast}(\mathbf{k},\omega)
\cdot
\hat{\mathbf{f}}(\mathbf{r},\omega).
\end{equation}

On the basis of the three unit vectors
$\mathbf{e}_{i}(\mathbf{k},\omega)$,
three (different) projection
kernels can be introduced,
\begin{equation}
\label{eq110}
\tensor{P}_{i}(\mathbf{r},\mathbf{r}',\omega)
=
\frac{1}{(2\pi)^{3}}\int \D^3k\,
\mathbf{e}_{i}(\mathbf{k},\omega)
\ot
\mathbf{e}_{i}^{\ast}(\mathbf{k},\omega)\,
e^{i\mathbf{k}\cdot(\mathbf{r}-\mathbf{r}')},
\end{equation}
provided that
$\sigma_{i}(\mathbf{k},\omega)$ $\!\neq$
$\!\sigma_{i'}(\mathbf{k},\omega)$ for
$i$ $\!\neq$ $i'$.

Let us consider, in particular,
isotropic media that in addition do not feature
optical activity in more detail.
In this case, the diagonal form of the tensor
$\tensor{\sigma}(\mathbf{k},\omega)$ reads
(see Ref.~\cite{MelroseBook})
\begin{equation}
\label{eq8-23-13}
\tensor{\sigma}(\mathbf{k},\omega)
=
\sigma_{\parallel}(k,\omega)
\frac{\mathbf{k}\ot\mathbf{k}}{k^2}
+
\sigma_{\perp}(k,\omega)
\left(\tensor{I}-\frac{\mathbf{k}\ot\mathbf{k}}{k^2}\right),
\end{equation}
i.e.,
$\sigma_1(k,\omega)$ $\!=$
$\!\sigma_\parallel(k,\omega)$
and $\sigma_2(k,\omega)$ $\!=$
$\!\sigma_3(k,\omega)$
$\!=$ $\!\sigma_\perp(k,\omega)$
$\!\neq$ $\!\sigma_\parallel(k,\omega)$,
which implies that $\tensor{K}(\mathbf{k},\omega)$,
Eq.~(\ref{eq120}), takes the form
\begin{equation}
\label{eq121}
\tensor{K}(\mathbf{k},\omega)
=
\sigma_{\parallel}^{1/2}(k,\omega)
\frac{\mathbf{k}\ot\mathbf{k}}{k^2}
+
\sigma_{\perp}^{1/2}(k,\omega)
\left(\tensor{I}-\frac{\mathbf{k}\ot\mathbf{k}}{k^2}\right).
\end{equation}
Thus, the well-known
longitudinal and transverse tensorial $\delta$-functions
\mbox{$\tensor{\Delta}_{\parallel}(\mathbf{r}-\mathbf{r}')$}
and
\mbox{$\tensor{\Delta}_{\perp}(\mathbf{r}-\mathbf{r}')$},
respectively, can be taken as projection kernels,
\begin{equation}
\label{NLB17}
\tensor{P}_{\parallel(\perp)}(\mathbf{r},\mathbf{r}',\omega)
=
\tensor{\Delta}_{\parallel(\perp)}(\mathbf{r}-\mathbf{r}'),
\end{equation}
which may be used to introduce, according to Eq.~(\ref{NL25d}), the
projective variables
\begin{equation}
\label{NLB17-1}
\hat{\mathbf{f}}_{\parallel(\perp)}(\mathbf{r},\omega)
=
\int \D^3s\,
\tensor{\Delta}_{\parallel(\perp)}(\mathbf{r}-\mathbf{s})
\cdot
\hat{\mathbf{f}}(\mathbf{s},\omega).
\end{equation}
\subsubsection{Unitarily equivalent formulation}
\label{sec4C-1}

As already pointed out in Sec.~\ref{sec2}, the integral
kernel $\tensor{K}(\mathbf{r},\mathbf{r}',\omega)$
in Eq.~(\ref{eq8-5}) is not uniquely determined by
Eq.~(\ref{eq8-7}), since any other kernel
$\tensor{K}{'}(\mathbf{r},\mathbf{r}',\omega)$
of the form (\ref{NL21-1}) [together with Eq.~(\ref{B3})]
is also an allowed kernel.
To illustrate
this for the isotropic medium under study, we first note
that Eq.~(\ref{eq8-23-13}) may be equivalently rewritten as
\begin{align}
\label{eq8-23-11}
\tensor{\sigma}(\mathbf{k},\omega)
&
=
\sigma_{\parallel}(k,\omega)
\tensor{I}
-
\mathbf{k}\times
\gamma(k,\omega)\tensor{I}
\times\mathbf{k},
\end{align}
where
\begin{equation}
\label{eq8-23-20}
\gamma(k,\omega)
=
[\sigma_{\perp}(k,\omega)
-
\sigma_{\parallel}(k,\omega)]
/k^2.
\end{equation}
Since
(for real $\omega$)
$\sigma_{\parallel}(k,\omega)$ and
$\sigma_{\perp}(k,\omega)$
are both real and positive
[in accordance with the requirement that
$\tensor{\sigma}(\mathbf{r},\mathbf{r}',\omega)$ be
the integral kernel of a positive definite operator],
$\gamma(k,\omega)$ is
real but its sign is not determined
by this requirement.
However, if $\gamma(k,\omega)$ is required here and below
to be positive throughout, then
\begin{equation}
\label{eq140}
\tensor{K}{'}(\mathbf{k},\omega)
=
\sigma_{\parallel}^{1/2}(k,\omega)\tensor{I}
\pm
\gamma^{1/2}(k,\omega)
\mathbf{k}\times\tensor{I}
\end{equation}
obeys the equation
\begin{equation}
\label{eq122}
\tensor{K}{'}(\mathbf{k},\omega)
\cdot\tensor{K}{^{\prime +}}(\mathbf{k},\omega)
=
\tensor{\sigma}(\mathbf{k},\omega).
\end{equation}
Moreover, it can be shown that
$\tensor{K}{'}(\mathbf{k},\omega)$ can be represented in
the form
\begin{equation}
\label{eq140-a}
\tensor{K}{'}(\mathbf{k},\omega)
=
\tensor{K}(\mathbf{k},\omega)
\cdot
\tensor{V}(\mathbf{k},\omega),
\end{equation}
with
\begin{multline}
\label{eq161}
\tensor{V}(\mathbf{k},\omega)=
\frac{\mathbf{k}\ot\mathbf{k}}{k^2}
+
\sigma_{\parallel}^{1/2}(k,\omega)
\sigma_{\perp}^{-1/2}(k,\omega)
\left(\tensor{I}-\frac{\mathbf{k}\ot\mathbf{k}}{k^2}\right)
\\
\pm
\gamma^{1/2}(k,\omega)
\sigma_{\perp}^{-1/2}(k,\omega)
\mathbf{k}\times \tensor{I}
\end{multline}
[$\tensor{V}{^{-1}}(\mathbf{k},\omega)$ $\!=$
$\tensor{V}{^{+}}(\mathbf{k},\omega)$].
Hence, $\tensor{K}{'}(\mathbf{k},\omega)$ also
yields, according to Eq.~(\ref{NLB12}), a valid integral kernel
$\tensor{K}{'}(\mathbf{r},\mathbf{r}',\omega)$,
\begin{equation}
\label{eq130}
\tensor{K}{'}(\mathbf{r},\mathbf{r}',\omega)
=
\frac{1}{(2\pi)^{3}}
\int \D^3k\,
\tensor{K}{'}(\mathbf{k},\omega)
e^{i\mathbf{k}\cdot(\mathbf{r}-\mathbf{r}')},
\end{equation}
which is related to the integral kernel
$\tensor{K}(\mathbf{r},\mathbf{r}',\omega)$
according to Eq.~(\ref{NL21-1}), where
\begin{equation}
\label{eq150}
\tensor{V}(\mathbf{r},\mathbf{r}',\omega)
=
\frac{1}{(2\pi)^{3}}
\int \D^3k\,
\tensor{V} (\mathbf{k},\omega)
e^{i\mathbf{k}\cdot(\mathbf{r}-\mathbf{r}')},
\end{equation}
with the associated operator being unitary.
We thus see that the two formulations of the theory based on
$\tensor{K}(\mathbf{r},\mathbf{r}',\omega)$
and $\tensor{K}{'}(\mathbf{r},\mathbf{r}',\omega)$,
respectively, are unitarily equivalent.
Note that $\tensor{K}{'}(\mathbf{k},\omega)$ $\!\neq$
$\tensor{K}{^{\prime +}}(\mathbf{k},\omega)$,
so that the operator associated with the integral
kernel $\tensor{K}{'}(\mathbf{r},\mathbf{r}',\omega)$
is non-Hermitian (as to be expected, see App.~\ref{AppB}).
Since the operators associated with
$\tensor{K}{'}(\mathbf{r},\mathbf{r}',\omega)$
[as well as $\tensor{V}(\mathbf{r},\mathbf{r}',\omega)$]
and $\tensor{P}_{\parallel(\perp)}(\mathbf{r},\mathbf{r}',\omega)$
commute, the same projectors
may be employed in the two
formulations of the theory
to introduce projective variables according to Eq.~(\ref{NLB17-1}).

\subsubsection{Local limit: magnetodielectric media}
\label{sec4C-2}

Now let us suppose that $\sigma_{\parallel}(k,\omega)$ and
$\gamma(k,\omega)$ in Eq.~(\ref{eq8-23-11})
are sufficiently slowly varying functions of
$k$,
with well-defined and unique
long-wavelength limits
$\lim_{k\to 0}$ $\!\sigma_{\parallel}(k,\omega)$
$\!=$ $\!\sigma_{\parallel}(\omega)$
$\!>0$
and
\mbox{$\lim_{k\to 0}$ $\!\gamma(k,\omega)$}
\mbox{$\!=$ $\!\gamma(\omega)$ $\!>0$},
so that they may be approximated by
these limits under the integral in Eq.~(\ref{eq8-20})
to obtain
\begin{multline}
\label{eq210}
\tensor{\sigma}(\mathbf{r},\mathbf{r}',\omega)
=
\sigma_{\parallel}(\omega)
\tensor{I}\delta(\mathbf{r-r'})
\\
-
\gamma(\omega)\Nabla\times[\tensor{I}
\delta(\mathbf{r-r'})]\times\Lnabla{'}.
\end{multline}
It should be pointed out
that in the limiting case
given by Eq.~(\ref{eq210})
the positive definiteness of
the operator associated with
$\tensor{\sigma}(\mathbf{r},\mathbf{r}',\omega)$
already implies that
$\gamma(\omega)$ must be positive,
$\gamma(\omega)\!>\!0$;
in the general case as given by Eq.~(\ref{eq8-20})
together with Eqs.~(\ref{eq8-23-11}) and (\ref{eq8-23-20}),
the positive definiteness
of
$\tensor{\sigma}(\mathbf{r},\mathbf{r}',\omega)$
does not automatically restrict
$\gamma(k,\omega)$ to positive values.

In order to see to what type of medium this
$\tensor{\sigma}(\mathbf{r},\mathbf{r}',\omega)$
corresponds, we have to find
from Eq.~(\ref{eq210}) the full conductivity tensor
$\tensor{Q}(\mathbf{r},\mathbf{r}',\omega)$, which is
uniquely possible since
Eqs.~(\ref{eq2}) and (\ref{eq3})
are Hilbert transforms of each other (cf.~Sec.~\ref{sec2}).
The full conductivity tensor
corresponding to Eq.~(\ref{eq210})
is thus of the form
\begin{multline}
\label{eq74}
\tensor{Q}(\mathbf{r},\mathbf{r}',\omega)
=
Q^{(1)}(\omega)
\tensor{I}\delta(\mathbf{r-r'})
\\
-
Q^{(2)}(\omega)
\Nabla\times[\tensor{I}
\delta(\mathbf{r-r'})]\times\Lnabla{'},
\end{multline}
where $Q^{(1)}(\omega)$ and
$Q^{(2)}(\omega)$ are (Fourier-transformed) response functions, both of
which are determined by their respective real parts
$\sigma_{\parallel}(\omega)$ and $\gamma(\omega)$.
Inserting
Eq.~(\ref{eq74}) into
Eq.~(\ref{eq8-1})
and comparing with
\begin{multline}
\label{eq230}
\fo{j}(\mathbf{r},\omega)
=
- i \varepsilon_{0} \omega \,
[\varepsilon(\omega)-1]\fo{E}(\mathbf{r},\omega)
\\
+
\kappa_{0}
\Nabla\times\{[1-\kappa(\omega)]\fo{B}(\mathbf{r},\omega)\}
+\fo{j}_{\mathrm N}(\mathbf{r},\omega)
\end{multline}
[$\fo{B}(\mathbf{r},\omega)$ $\!=$ $\!(i\omega)^{-1}\Nabla\times
\fo{E}(\mathbf{r},\omega)$],
which is the well-known description of
a locally responding (homogeneous)
magnetodielectric medium, we can
make the identifications
\begin{equation}
\label{eq76}
Q^{(1)}(\omega)
=
- i \varepsilon_{0} \omega \,[\varepsilon(\omega)-1]
\end{equation}
and
\begin{equation}
\label{eq77}
Q^{(2)}(\omega)
=
-i\kappa_{0}[1-\kappa(\omega)]/ \omega,
\end{equation}
where
$\varepsilon(\omega)$ is the permittivity and
$\mu(\omega)\!=\!\kappa^{-1}(\omega)$ the
(paramagnetic) permeability of the medium
($\mu_{0}\!=\!\kappa_{0}^{-1}$).
For real $\omega$, we thus obtain
\begin{equation}
\label{eq700}
\sigma_{\parallel}(\omega)
=
\varepsilon_{0} \omega \,\Im \varepsilon(\omega)
\end{equation}
and
\begin{equation}
\label{eq710}
\gamma(\omega)
=
-\kappa_{0} \Im \kappa(\omega)/\omega.
\end{equation}
Note that, because of
$\gamma(\omega)$ $\!>$ $\!0$,
from Eq.~(\ref{eq710}) it follows that
$\Im\kappa(\omega)$ $\!<$ $\!0$ for $\omega\!>\!0$,
from which it can be shown that
$\mu(\omega\!\to\!0)$ $\!>$ $\!1$.

At first glance, one
might believe (erroneously) that
not only paramagnetic
[$\mu(\omega\!\to\!0)$ $\!>$ $\!1$]
but also diamagnetic
[$\mu(\omega\!\to\!0)$ $\!<$ $\!1$]
features of a
medium (or the combined effect of both)
can be consistently described
by means of the magnetic permeability $\mu(\omega)$
which is included, as seen above,
in the basic linear-response
constitutive relation (\ref{eq8-1}).
However, since diamagnetism is basically
a nonlinear effect (as the underlying
microscopic Hamiltonian is quadratic
in the magnetic induction field), it
is beyond the scope of linear response theory.
If it is desired to include diamagnetic media in the framework of linear
electrodynamics nevertheless, one can regard the magnetic field on
which the diamagnetic susceptibility
depends as being (the mean value of) an
externally controlled field independent of the
dynamical variables. 
Note that the
Onsager reciprocity theorem needs to
be stated in its generalized form in this case, see
Refs.~\cite{KuboNonEqStat,MelroseBook}.
For a more satisfactory account of diamagnetic media, one should,
however, resort to a non-linear response formalism, or to a more
microscopic theory.

An obvious solution to
Eq.~(\ref{eq8-7}) with $\tensor{\sigma}(\mathbf{r},\mathbf{r}',\omega)$
given by Eq.~(\ref{eq210}) is provided by
\begin{equation}
\label{eq181}
\tensor{K}{'}(\mathbf{r},\mathbf{r}',\omega)
=
\sigma_{\parallel}^{1/2}(\omega)
\tensor{I}\delta(\mathbf{r}-\mathbf{r}')
\mp i
\gamma^{1/2}(\omega)
\Nabla \times \tensor{I}\delta(\mathbf{r}-\mathbf{r}'),
\end{equation}
which corresponds to the kernel (\ref{eq130})
[together with Eq.~(\ref{eq140})] when,
for an isotropic medium,
$\sigma_{\parallel}(k,\omega)$ and
$\gamma(k,\omega)$
are approximated by
$\sigma_{\parallel}(\omega)$ and $\gamma(\omega)$,
respectively.
The kernel (\ref{eq130}) [together with Eq.~(\ref{eq140})]
fits well here since it
depends in a particularly
simple way on those quantities that we have assumed to approach
well-defined limits in the derivation
that led to Eq.~(\ref{eq210}), a property
which can be attributed to the responsible
transformation (\ref{eq150}) [together with (\ref{eq161})].
In contrast, the kernel obtained directly from
Eq.~(\ref{NLB12}) [together with Eq.~(\ref{eq121})],
by first eliminating
$\sigma_{\perp}(k,\omega)$
by means of Eq.~(\ref{eq8-23-20}) and then
approximating
$\sigma_{\parallel}(k,\omega)\!\mapsto\!\sigma_{\parallel}(\omega)$
and $\gamma(k,\omega)\mapsto\!\gamma(\omega)$,
does not provide an
alternative to Eq.~(\ref{eq181}), as it
does not lead to Eq.~(\ref{eq210}) when
inserted in Eq.~(\ref{eq8-7}); it corresponds to
a different medium.
Correspondingly, the
kernel $\tensor{V}(\mathbf{r,r'},\omega)$
obtained by expressing Eq.~(\ref{eq150}) [with Eq.~(\ref{eq161})]
in terms of
$\sigma_{\parallel}(k,\omega)$
and
$\gamma(k,\omega)$
and
then approximating them by $\sigma_{\parallel}(\omega)$
and $\gamma(\omega)$, respectively, fails to be unitary.
Substituting for $\tensor{K}(\mathbf{r},\mathbf{r}',\omega)$
in Eq.~(\ref{eq8-5}) $\tensor{K}{'}(\mathbf{r},\mathbf{r}',\omega)$
as given by Eq.~(\ref{eq181}) [together with
Eqs.~(\ref{eq700}) and (\ref{eq710})],
we may explicitly express the noise current density in terms of
the bosonic dynamical variables to obtain
\begin{multline}
\label{eq220}
\fo{j}_{\mathrm N}(\mathbf{r},\omega)
=
\left(\frac{\hbar\varepsilon_{0}}{\pi}\right)^{\frac{1}{2}}
\sqrt{\omega^2 \Im \varepsilon(\omega)}\,
\hat{\mathbf{f}}(\mathbf{r},\omega)
\\
\mp i
\left(\frac{\hbar \kappa_{0}}{\pi}\right)^{\frac{1}{2}}
\Nabla\times
\bigl[
\sqrt{-\Im \kappa(\omega)}\,
\hat{\mathbf{f}}(\mathbf{r},\omega)
\bigr].
\end{multline}
Since the operators associated with
the projection kernels (\ref{NLB17})
commute with the operators associated with
Eqs.~(\ref{eq181}) and (\ref{eq210}),
one may introduce the
projective variables
$\hat{\mathbf{f}}_{\parallel(\perp)}(\mathbf{r},\omega)$
defined by Eq.~(\ref{NLB17-1}), which corresponds
to a decomposition of $\fo{j}_{\mathrm N}(\mathbf{r},\omega)$
into longitudinal and transverse parts,
\begin{equation}
\label{eq300}
\fo{j}_{\mathrm N}(\mathbf{r},\omega)
=
\fo{j}_{\mathrm N \parallel}(\mathbf{r},\omega)+
\fo{j}_{\mathrm N \perp}(\mathbf{r},\omega),
\end{equation}
where
\begin{equation}
\label{eq301}
\fo{j}_{\mathrm N \parallel}(\mathbf{r},\omega)
=
\left(\frac{\hbar\varepsilon_{0}}{\pi}\right)^{\frac{1}{2}}
\sqrt{\omega^2 \Im \varepsilon(\omega)}\,
\hat{\mathbf{f}}_{\parallel}(\mathbf{r},\omega),
\end{equation}
\begin{multline}
\label{eq302}
\fo{j}_{\mathrm N \perp}(\mathbf{r},\omega)
=
\left(\frac{\hbar\varepsilon_{0}}{\pi}\right)^{\frac{1}{2}}
\sqrt{\omega^2 \Im \varepsilon(\omega)}\,
\hat{\mathbf{f}}_{\perp}(\mathbf{r},\omega)
\\
\mp i
\left(\frac{\hbar \kappa_{0}}{\pi}\right)^{\frac{1}{2}}
\Nabla\times
\bigl[
\sqrt{-\Im \kappa(\omega)}\,
\hat{\mathbf{f}}
_{\perp}(\mathbf{r},\omega)
\bigr].
\end{multline}
Making use of Eq.~(\ref{NL25d}) and identifying therein
the projection kernels $\tensor{P}_\lambda(\mathbf{r},
\mathbf{r}',\omega)$ with
$\tensor{\Delta}_{\parallel(\perp)}(\mathbf{r-r'})$, 
one may then proceed as described in Sec.~\ref{sec3} and regard
the projective variables
$\hat{\mathbf{f}}_{\parallel}(\mathbf{r},\omega)$ and
$\hat{\mathbf{f}}_{\perp}(\mathbf{r},\omega)$
as being two independent sets of bosonic
variables.

Let us briefly make contact with the quantization scheme
described in Refs.~\cite{KnoellL002001,HoTD102003},
where $\fo{j}_{\mathrm N}(\mathbf{r},\omega)$
is decomposed according to
\begin{equation}
\label{eq306}
\fo{j}_{\mathrm N}(\mathbf{r},\omega)
=
\fo{J}_{\mathrm N e}(\mathbf{r},\omega)+
\fo{J}_{\mathrm N m}(\mathbf{r},\omega),
\end{equation}
with
\begin{align}
\label{eq307}
&\fo{J}_{\mathrm N e}(\mathbf{r},\omega)
=
\left(\frac{\hbar\varepsilon_{0}}{\pi}\right)^{\frac{1}{2}}
\sqrt{\omega^2 \Im \varepsilon(\omega)}\,
\hat{\mathbf{f}}_{e}(\mathbf{r},\omega),
\\
\label{eq308}
&\fo{J}_{\mathrm N m}(\mathbf{r},\omega)
=
\mp i
\left(\frac{\hbar \kappa_{0}}{\pi}\right)^{\frac{1}{2}}
\Nabla\times
\bigl[
\sqrt{-\Im \kappa(\omega)}\,
\hat{\mathbf{f}}_{m}(\mathbf{r},\omega)
\bigr].
\end{align}
The connection between
Eqs.~(\ref{eq300})--(\ref{eq302})
and Eqs.~(\ref{eq306})--(\ref{eq308})
is given by a gauge transformation
(cf.~Sec.~\ref{sec3}),
which effectively redistributes
the first term of Eq.~(\ref{eq302}).
It is not difficult to prove that the 
total noise current as given by Eq.~(\ref{eq306})
satisfies the correct commutation
relation (\ref{eq8-6}) [with $\tensor{\sigma}(\mathbf{r},
\mathbf{r}',\omega)$ from Eq.~(\ref{eq210}) together with
Eqs.~(\ref{eq700}) and (\ref{eq710})]
if $\hat{\mathbf{f}}_{e}(\mathbf{r},\omega)$ and
$\hat{\mathbf{f}}_{m}(\mathbf{r},\omega)$
are regarded
as two independent sets of bosonic variables.
Since $\fo{J}_{\mathrm N e}(\mathbf{r},\omega)$ and
$\fo{J}_{\mathrm N m}(\mathbf{r},\omega)$ can be
linearly related to
$\fo{j}_{\mathrm N \parallel}(\mathbf{r},\omega)$ and
$\fo{j}_{\mathrm N \perp}(\mathbf{r},\omega)$ and thus to
$\fo{j}_{\mathrm N}(\mathbf{r},\omega)$,
the variables
$\hat{\mathbf{f}}_{e}(\mathbf{r},\omega)$ and
$\hat{\mathbf{f}}_{m}(\mathbf{r},\omega)$
may be viewed as projective variables,
in the sense outlined at the end of Sec.\ref{sec3}.
Since Eq.~(\ref{eq306}) [with Eqs.~(\ref{eq307}) and (\ref{eq308})]
is a separation of the noise current into a part
attributed to a dielectric polarization and a part attributed to a
(paramagnetic) magnetization, 
the quantization scheme based on Eqs.~(\ref{eq306})--(\ref{eq308}),
with $\hat{\mathbf{f}}_{e}(\mathbf{r},\omega)$ and
$\hat{\mathbf{f}}_{m}(\mathbf{r},\omega)$ being bosonic variables,
may be thought of as following from the general quantization scheme
in the case where magneto-electric crossing effects
can be a priori excluded from consideration.

\subsubsection{Local limit: other kinds of media}
\label{sec4C-3}

The transition to the local limit is not a unique procedure in general.
Various kinds of locally responding
(homogeneous) media, including non-isotropic ones,
may therefore be derived as limiting cases from
Eq.~(\ref{eq8-20}).
To illustrate this, let us represent
$\tensor{\sigma}(\mathbf{k},\omega)$ as given in
Eq.~(\ref{eq8-21}) in a different orthonormal
basis, where the new expansion will be non-diagonal in general,
\begin{equation}
\label{eq181-1}
\tensor{\sigma}(\mathbf{k},\omega)
=\sum_{i,j=1}^{3}
\tilde{\sigma}_{ij}(\mathbf{k},\omega)\,
\mathbf{\tilde{e}}_{i}(\mathbf{k},\omega)
\ot
\mathbf{\tilde{e}}_{j}^{\ast}(\mathbf{k},\omega).
\end{equation}
The new basis vectors
$\mathbf{\tilde{e}}_{i}(\mathbf{k},\omega)$
are related to the ones appearing in Eq.~(\ref{eq8-21})
by a unitary transformation,
\begin{align}
\label{eq182}
\mathbf{\tilde{e}}_{i}(\mathbf{k},\omega)
=\sum_{k=1}^{3}
U_{ik}(\mathbf{k},\omega) \mathbf{e}_{k}(\mathbf{k},\omega),
\\
\label{eq183}
U_{ik}(\mathbf{k},\omega)
=
\mathbf{\tilde{e}}_{i}(\mathbf{k},\omega)
\cdot
\mathbf{e}_{k}^{\ast}(\mathbf{k},\omega).
\end{align}
We may always choose the
$\mathbf{\tilde{e}}_{i}(\mathbf{k},\omega)$ so
that they are independent of $\mathbf{k}$,
$\mathbf{\tilde{e}}_{i}(\mathbf{k},\omega)\mapsto
\mathbf{\tilde{e}}_{i}(\omega)$.
If this choice can be made such that
the new expansion coefficients,
\begin{equation}
\label{eq184}
\tilde{\sigma}_{ij}(\mathbf{k},\omega)
=\sum_{k,l=1}^{3}
U_{ik}^{\ast}(\mathbf{k},\omega)
\sigma_{kl}(\mathbf{k},\omega)
U_{jl}(\mathbf{k},\omega),
\end{equation}
may be approximately replaced under the
$\mathbf{k}$-integral according to
\begin{equation}
\label{eq185}
\tilde{\sigma}_{ij}(\mathbf{k},\omega)
\mapsto
\tilde{\sigma}_{ij}(\mathbf{k}\!\to\!0,\omega)\equiv
\tilde{\sigma}_{ij}(\omega)
\end{equation}
when Eq.~(\ref{eq181-1}) is inserted in Eq.~(\ref{eq8-20}),
then in this way
the type of locally responding (homogeneous)
anisotropic medium defined by
Eq.~(\ref{NLB1}) is recovered. [Equation (\ref{NLB1a}) is then obtained by
diagonalizing $\tilde{\sigma}_{ij}(\omega)$
by means of yet another ($\mathbf{k}$-independent) unitary transformation.]
Similarly, if the approximation (\ref{eq185}) is generalized
to include further terms of an (assumed) expansion of
$\tilde{\sigma}_{ij}(\mathbf{k},\omega)$ at $\mathbf{k}\!=\!0$, then
quasi-local approximations of Eq.~(\ref{eq8-20}) are
generated, by inserting the truncated expansion into
Eq.~(\ref{eq8-20}) and integrating term by term to
yield a linear combination of various derivatives of
$\delta$-functions.
In pursuing such approximation procedures---whose validity
is to be examined in each case and which depends crucially on the
choice of the transformation (\ref{eq182}), (\ref{eq183})
(i.e., of the new basis vectors)---%
it must be kept in mind that
any approximate form of
$\tensor{\sigma}(\mathbf{r},\mathbf{r'},\omega)$ so derived
has to conform to all the general requirements on
$\tensor{Q}(\mathbf{r},\mathbf{r'},\omega)$.
In fact, the most general kind of linear medium discussed in
terms of local constitutive relations in the literature,
the so-called bi-anisotropic medium (see, e.g.,
Ref.~\cite{ChewBook}), may be viewed in this way
as a quasi-local approximation of
$\tensor{\sigma}(\mathbf{r},\mathbf{r'},\omega)$
[and thereby of $\tensor{Q}(\mathbf{r},\mathbf{r'},\omega)$]
that incorporates derivatives of $\delta$-functions up to
the second order.
%
%
\subsection{Spatially dispersive inhomogeneous media}
\label{sec4A}
As already mentioned, knowledge of
the medium properties, i.e., of
$
\tensor{\sigma}(\mathbf{r},\mathbf{r}',\omega)
$,
is required in order to solve the eigenvalue
problem (\ref{NL16}) and perform explicitly the quantization
the medium-assisted electromagnetic field%
---a task which, in general, cannot be accomplished in closed form.
Nevertheless, to provide some analytical insight into
the problem, let us consider media that combine
the features of the media considered in
Secs.~\ref{sec4B} and \ref{sec4C} in an
approximate fashion.

\subsubsection{Model}
\label{sec4A1}

We assume that the medium permits one to clearly
distinguish between the length scales associated with
spatial dispersion and inhomogeneity,
with the former scale being sufficiently
small as compared with the latter one.
In this case, the medium can be regarded as having locally
the properties of bulk material, and
$\tensor{\sigma}(\mathbf{r},\mathbf{r}',\omega)$
may be approximated as
\begin{multline}
\label{eq8-18}
\tensor{\sigma}(\mathbf{r},\mathbf{r}',\omega)
=
\frac{1}{\Omega}\sum_{\mathbf{L}}
\sum_{i=1}^{3}
\sum_{\mathbf{k}}
\sigma_{i\mathbf{L}\mathbf{k}}(\omega)
e^{i\mathbf{k}\cdot(\mathbf{r}-\mathbf{r}')}
\\ \times
\theta_\mathbf{L}(\mathbf{r})
\theta_\mathbf{L}(\mathbf{r}')\,
\mathbf{e}_{i\mathbf{L}\mathbf{k}}(\omega)
\ot
\mathbf{e}_{i\mathbf{L}\mathbf{k}}^{\ast}(\omega),
\end{multline}
from which the eigenfunctions of the associated operator
are seen to be
\begin{equation}
\label{eq8-19}
\mathbf{F}_{i\mathbf{L}\mathbf{k}}(\mathbf{r},\omega)
=
\Omega^{-1/2}
\theta_\mathbf{L}(\mathbf{r})\,
e^{i\mathbf{k}\cdot\mathbf{r}}
\mathbf{e}_{i\mathbf{L}\mathbf{k}}(\omega).
\end{equation}
Here, the medium is thought of as being divided into
unit cells of volume $\Omega$
which form a Bravais-type lattice,
the cut-off function $\theta_\mathbf{L}(\mathbf{r})$
is unity if $\mathbf{r}$ is in the cell
of lattice vector
$\mathbf{L}$ and zero otherwise,
$\mathbf{e}_{i\mathbf{L}\mathbf{k}}(\omega)$
are, for chosen $\mathbf{L}$, $\mathbf{k}$, and $\omega$, a
triplet ($i$ $\!=$ $\!1,2,3$)
of orthogonal unit vectors,
and the wave vector $\mathbf{k}$ runs
over the reciprocal lattice.
Note that, for each cell $\mathbf{L}$,
\begin{equation}
\label{eq100}
\tensor{\sigma}_{\mathbf{L}\mathbf{k}}(\omega)
=
\sum_{i=1}^{3}
\sigma_{i\mathbf{L\mathbf{k}}}(\omega)
\mathbf{e}_{i\mathbf{L}\mathbf{k}}(\omega)
\ot
\mathbf{e}_{i\mathbf{L}\mathbf{k}}^\ast(\omega)
\end{equation}
corresponds to the diagonal form in the
$(\mathbf{k},\omega)$ domain
of $\tensor{\sigma}(\mathbf{r},\mathbf{r}',\omega)$
for bulk material [cf.~Eq.~(\ref{eq8-21})].

The main features of
$\tensor{\sigma}(\mathbf{r},\mathbf{r}',\omega)$
as given in Eq.~(\ref{eq8-18}) can be summarized as follows.
(i)~$\tensor{\sigma}(\mathbf{r},\mathbf{r}',\omega)$
is zero whenever $\mathbf{r}$ and $\mathbf{r}'$ are not
in the same cell, so that
$\Omega^{1/3}$ determines the length scale
on which spatial dispersion is at most observed.
(ii)~The dependence on $\mathbf{L}$ of
$\tensor{\sigma}_{\mathbf{L}\mathbf{k}}(\omega)$
[Eq.~(\ref{eq100})] for an inhomogeneous medium introduces an
$\mathbf{L}$-dependence
into Eq.~(\ref{eq8-18}) which should be sufficiently weak,
so that noticeable violations of the translational
invariance of $\tensor{\sigma}(\mathbf{r},\mathbf{r}',\omega)$
may occur only on a length scale that is
large compared with $\Omega^{1/3}$.
Needless to say that
the main features do not essentially change if
$\theta_\mathbf{L}(\mathbf{r})$ is replaced by
another---but qualitatively similar---cut-off function.

Let us denote
by $\mathbf{L}(\mathbf{r})$ the particular
lattice vector whose cell contains the point $\mathbf{r}$, so that
$\mathbf{L}(\mathbf{r})$ plays the role of a
coarse-grained position variable.
With the notations
$\theta_\mathbf{L(\mathbf{r})}(\mathbf{r}')\mapsto
\theta[\mathbf{L}(\mathbf{r}),\mathbf{r}']$
and
$\tensor{\sigma}_{\mathbf{L}(\mathbf{r})\mathbf{k}}(\omega)
\mapsto \tensor{\sigma}_\mathbf{k}[\mathbf{L}(\mathbf{r}),\omega]$,
Eq.~(\ref{eq8-18}) together with Eq.~(\ref{eq100})
can be rewritten as
\begin{equation}
\label{eq8-30}
\tensor{\sigma}(\mathbf{r},\mathbf{r}',\omega)
=
\theta[\mathbf{L}(\mathbf{r}'),\mathbf{r}]
\tensor{\sigma}[\mathbf{L}(\mathbf{r'}),\mathbf{r-r'},\omega],
\end{equation}
with
\begin{equation}
\label{eq8-30-1}
\tensor{\sigma}[\mathbf{L}(\mathbf{r'}),\mathbf{r-r'},\omega]
=
\frac{1}{\Omega}
\sum_{\mathbf{k}}
\tensor{\sigma}_\mathbf{k}
[\mathbf{L}(\mathbf{r}'),\omega]
e^{i\mathbf{k}\cdot(\mathbf{r}-\mathbf{r}')}.
\end{equation}
Note that for arbitrary
(continuous) values $\mathbf{s}$, the function
$\theta(\mathbf{s},\mathbf{r})$
can be regarded as being symmetric.
Using Eq.~(\ref{eq8-19}), we find that
Eq.~(\ref{NL21}) takes the form
\begin{multline}
\label{eq101}
\tensor{K}(\mathbf{r},\mathbf{r}',\omega)
=
\frac{\theta[\mathbf{L}(\mathbf{r}'),\mathbf{r}]}{\Omega}
\sum_{\mathbf{k}}
\tensor{K}_\mathbf{k}[\mathbf{L}(\mathbf{r}'),\omega]
e^{i\mathbf{k}\cdot(\mathbf{r}-\mathbf{r}')},
\end{multline}
where
\begin{equation}
\label{eq102}
\tensor{K}_\mathbf{k}[\mathbf{L}(\mathbf{r}),\omega]
=
\sum_{i=1}^{3}
\sigma_{i\mathbf{k}}^{1/2}[\mathbf{L}(\mathbf{r}),\omega]\,
\mathbf{e}_{i\mathbf{k}}[\mathbf{L}(\mathbf{r}),\omega]
\ot
\mathbf{e}_{i\mathbf{k}}^{\ast}[\mathbf{L}(\mathbf{r}),\omega]
\end{equation}
$\{
\sigma_{i\mathbf{L}
(\mathbf{r})
\mathbf{k}}(\omega)
\mapsto
\sigma_{i\mathbf{k}}[\mathbf{L}(\mathbf{r}),\omega]$,
$\mathbf{e}_{i\mathbf{L}
%
(\mathbf{r})
\mathbf{k}}(\omega)
\mapsto \mathbf{e}_{i\mathbf{k}}[\mathbf{L}(\mathbf{r}),\omega]
\}$.

It can be shown that
Eq.~(\ref{eq8-30}) [with Eq.~(\ref{eq8-30-1})]
indeed contains (and, in a
sense, interpolates) the two limiting cases studied in
Secs.~\ref{sec4B} and \ref{sec4C}. For the proof, we observe
that in the case of negligible spatial dispersion, the
cell size can be shrunk to zero, $\Omega\!\to\!0$, so that
the lattice vectors take on continuous values,
$\mathbf{L}(\mathbf{r})\!\to\!\mathbf{r}$.
As the lattice becomes finer and finer,
the reciprocal lattice becomes more and more
coarse, and, for
$\mathbf{r}$ and $\mathbf{r'}$ unequal but
in the same cell, all the points of the reciprocal lattice
with $\mathbf{k}$ $\!\neq$ $\!0$ give rise to rapidly
oscillating terms in Eq.~(\ref{eq8-30-1}).
In the limit $\Omega\!\to\!0$, these terms
oscillate infinitely rapidly and
average to zero (when applying the operator
associated with Eq.~(\ref{eq8-30}) [with Eq.~(\ref{eq8-30-1})]
to any reasonable function),
so that they may be set equal to zero.
Taking also into account that
$\theta[\mathbf{L}(\mathbf{r}'),\mathbf{r}]/\Omega
\to\delta(\mathbf{r}-\mathbf{r}')$ in this limit,
we see that
Eq.~(\ref{eq8-30}) [with Eq.~(\ref{eq8-30-1})]
indeed approaches Eq.~(\ref{NLB1}) for vanishing spatial dispersion
[note the correspondences
$\sigma_{i\mathbf{k}=0}[\mathbf{L}(\mathbf{r})$ $\!=$
$\!\mathbf{r},\omega]$ $\!=$
$\!\sigma_{i}(\mathbf{r},\omega)$
and
$\mathbf{e}_{i\mathbf{k}=0}[\mathbf{L}(\mathbf{r})$ $\!=$
$\!\mathbf{r},\omega]$ $\!=$
$\!\mathbf{e}_{i}(\mathbf{r},\omega)$
].

On the other hand, in the limiting case of an infinitely extended homogeneous
medium, there is no $\mathbf{L}$-dependence
of the medium properties so that
we are free to increase the cell size indefinitely,
$\Omega\to\infty$. Consequently, we may let
$\theta[\mathbf{L}(\mathbf{r}'),\mathbf{r}]\to 1$
in Eq.~(\ref{eq8-30}) and
$\tensor{\sigma}_\mathbf{k}[\mathbf{L}(\mathbf{r}),\omega]\to
\tensor{\sigma}(\mathbf{k},\omega)$,
$\Omega^{-1}\sum_{\mathbf{k}}
\to(2\pi)^{-3}\int\D^3k$ in
Eq.~(\ref{eq8-30-1}),
which reveals that
Eq.~(\ref{eq8-30}) [with Eq.~(\ref{eq8-30-1})]
approaches Eq.~(\ref{eq8-20}) as expected.


\subsubsection{Magnetodielectric media}
\label{sec4A2}

To quantize the electromagnetic field in an inhomogeneous
magnetodielectric medium specified
in terms of
$\varepsilon(\mathbf{r},\omega)$ and $\kappa(\mathbf{r},\omega)$
$\!=$ $\!\mu^{-1}(\mathbf{r},\omega)$,
let us consider a medium that is both sufficiently
weakly inhomogeneous and sufficiently weakly spatially
dispersive, so that $\Omega$ in Eq.~(\ref{eq8-30-1})
can be chosen on a scale intermediate between the scales
of spatial dispersion and inhomogeneity.
We may then approximately let $\mathbf{L}(\mathbf{r})$
be a continuous variable,
\mbox{$\mathbf{L}(\mathbf{r})\!\to\!\mathbf{r}$}
in Eq.~(\ref{eq8-30}),
and yet, at the same time,
approximately treat the $\mathbf{k}$-sum
in Eq.~(\ref{eq8-30-1})
as an integral, so that
Eq.~(\ref{eq8-30})
[with Eq.~(\ref{eq8-30-1})]
approximates to
[$\theta[\mathbf{L}(\mathbf{r}'),\mathbf{r}]
\to \theta(\mathbf{r}',\mathbf{r})$]
\begin{equation}
\label{eq400}
\tensor{\sigma}(\mathbf{r},\mathbf{r}',\omega)
=
\frac{\theta(\mathbf{r}',\mathbf{r})}
{(2\pi)^{3}}\int \D^3k\,
\tensor{\sigma}(\mathbf{r}',\mathbf{k},\omega)
e^{i\mathbf{k}\cdot(\mathbf{r}-\mathbf{r}')}.
\end{equation}
For a medium that is locally of the type
described by Eq.~(\ref{eq8-23-11}), we may set
\begin{equation}
\label{eq410}
\tensor{\sigma}(\mathbf{r},\mathbf{k},\omega)
=
\sigma_{\parallel}(\mathbf{r},k,\omega)\tensor{I}
-
\mathbf{k}\times \gamma(\mathbf{r},k,\omega)
\tensor{I}\times\mathbf{k},
\end{equation}
where
\begin{equation}
\label{eq440}
\gamma(\mathbf{r},k,\omega)
=
[\sigma_{\perp}(\mathbf{r},k,\omega)
-
\sigma_{\parallel}(\mathbf{r},k,\omega)]/k^2
> 0.
\end{equation}
Assuming that in the $\mathbf{k}$-integral in Eq.~(\ref{eq400}),
$\sigma_{\parallel}(\mathbf{r},k,\omega)$ and
$\gamma(\mathbf{r},k,\omega)$
may be approximated, respectively, by
well-defined (and unique) long-wavelength limits
$\sigma_{\parallel}(\mathbf{r},\omega)$ $\!=$
$\lim_{k\to 0}$
$\!\sigma_{\parallel}(\mathbf{r},k,\omega)$
and
$\gamma(\mathbf{r},\omega)$ $\!=$
$\lim_{k\to 0}\gamma(\mathbf{r},k,\omega)$,
the cut-off function
$\theta(\mathbf{r}',\mathbf{r})$
has---due to the rapid oscillations of the exponential
for large $|\mathbf{r}-\mathbf{r}'|$---no effect
[with regard to an
application of the operator associated with
$\tensor{\sigma}(\mathbf{r},\mathbf{r}',\omega)$ from
Eq.~(\ref{eq400})]
and can be dropped, and we obtain, as a
generalization of Eq.~(\ref{eq210}),
\begin{multline}
\label{eq490}
\tensor{\sigma}(\mathbf{r},\mathbf{r}',\omega)
=
\sigma_{\parallel}(\mathbf{r'},\omega)
\tensor{I}\delta(\mathbf{r}-\mathbf{r}')
\\
-
\Nabla\times[
\gamma(\mathbf{r'},\omega)
\tensor{I}\delta(\mathbf{r}-\mathbf{r}')]\times\Lnabla{'}.
\end{multline}
With the identifications
\begin{align}
\label{eq600}
&\sigma_{\parallel}(\mathbf{r},\omega)
=
\varepsilon_{0} \omega \,\Im \varepsilon(\mathbf{r},\omega),
\\
\label{eq610}
&\gamma(\mathbf{r},\omega)
=
-\kappa_{0} \Im \kappa(\mathbf{r},\omega)/\omega
\end{align}
[cf. Eqs.~(\ref{eq700}) and (\ref{eq710})],
Eqs.~(\ref{eq74}) and (\ref{eq230})
generalize to
\begin{multline}
\label{eq620}
\tensor{Q}(\mathbf{r,r'},\omega)
=
- i \varepsilon_{0} \omega
\,[\varepsilon(\mathbf{r'},\omega)-1]
\tensor{I}\delta(\mathbf{r-r'})
\\
-
\frac{1}{i \omega}
\Nabla\times\{
\kappa_{0}[1-\kappa(\mathbf{r'},\omega)]
\tensor{I}\delta(\mathbf{r-r'})
\}\times\Lnabla{'}
\end{multline}
and
\begin{multline}
\label{eq630}
\fo{j}(\mathbf{r},\omega)
=
- i \varepsilon_{0} \omega \,
[\varepsilon(\mathbf{r},\omega)-1]
\fo{E}(\mathbf{r},\omega)
\\
+
\kappa_{0}
\Nabla\times\{[1-\kappa(\mathbf{r},\omega)]
\fo{B}(\mathbf{r},\omega)\}
+\fon{j}_{\mathrm N}(\mathbf{r},\omega),
\end{multline}
respectively.

Unfortunately, Eq.~(\ref{eq181}) does not generalize to
\begin{multline}
\label{eq450}
\tensor{K}{'}(\mathbf{r},\mathbf{r}',\omega)
=
\sigma_{\parallel}^{1/2}(\mathbf{r}',\omega)
\tensor{I}\delta(\mathbf{r}-\mathbf{r}')
\\
\mp i
\gamma^{1/2}(\mathbf{r}',\omega)
\Nabla\times\tensor{I}\delta(\mathbf{r}-\mathbf{r}')
\quad \mbox{(wrong!)},
\end{multline}
as could have been suspected.
Indeed, straightforward calculation shows that,
for spatially varying permittivity and permeability,
the kernel (\ref{eq450}) does not solve
Eq.~(\ref{eq8-7})
[with $\tensor{\sigma}(\mathbf{r},\mathbf{r}',\omega)$
as given in Eq.~(\ref{eq490})],
which implies that $\fo{j}_{\mathrm N}(\mathbf{r},\omega)$ cannot be
related to the variables $\hat{\mathbf{f}}(\mathbf{r},\omega)$
as in Eq.~(\ref{eq220}), with $\varepsilon(\omega)$
and $\kappa(\omega)$ being simply replaced with their
inhomogeneous counterparts
$\varepsilon(\mathbf{r},\omega)$ and
$\kappa(\mathbf{r},\omega)$, respectively.
In order to obtain an explicit expression for
the kernel $\tensor{K}(\mathbf{r},\mathbf{r}',\omega)$
required in Eq.~(\ref{eq8-5}),
one has instead to return to Eq.~(\ref{eq8-7}) and solve it
with $\tensor{\sigma}(\mathbf{r},\mathbf{r}',\omega)$
from Eq.~(\ref{eq490})---a problem that is, however, very
difficult to solve in general.
Although this does not at all limit the practical
applicability of the theory
[since all one typically has to know about
$\tensor{K}(\mathbf{r},\mathbf{r}',\omega)$
is that it satisfies its
defining equation (\ref{eq8-7})
for the chosen conductivity (\ref{eq620})],
it may be useful to have at
hand, at least an approximate form for weak inhomogeneity,
such as (App.~\ref{AppG})
\begin{multline}
\label{eq500}
\tensor{K}(\mathbf{r},\mathbf{r}',\omega)
=
\sigma_{\parallel}^{1/2}(\omega)
\tensor{I}\delta(\mathbf{r}-\mathbf{r}')
\mp i
\gamma^{1/2}(\omega)
\Nabla \times \tensor{I}\delta(\mathbf{r}-\mathbf{r}')
\\
+
{\textstyle\frac{1}{2}}
[
\sigma_{\parallel}(\mathbf{r},\omega)
-
\sigma_{\parallel}(\omega)]
\tensor{M}_{0}(\mathbf{r},\mathbf{r}',\omega)
\\
+
{\textstyle\frac{1}{2}}
\Nabla\times
\bigl\{
[\gamma(\mathbf{r},\omega)-\gamma(\omega)]
\Nabla\times\tensor{M}_{0}(\mathbf{r},\mathbf{r}',\omega)
\bigr\},
\end{multline}
where
\begin{align}
\label{G9}
\tensor{M}_{0}(\mathbf{r},\mathbf{r}',\omega)
&
=\sigma_{\parallel}^{-1/2}(\omega)
\tensor{I}\delta(\mathbf{r}-\mathbf{r}')
\nonumber\\
&
\pm i
\gamma^{-1/2}(\omega)
\Nabla\times m_{0}(\mathbf{r},\mathbf{r}',\omega)\tensor{I}
\nonumber\\
&
+
\sigma_{\parallel}^{-1/2}(\omega)
\Nabla\times m_{0}(\mathbf{r},\mathbf{r}',\omega)
\tensor{I}\times\Lnabla{'},
\end{align}
with
$m_{0}(\mathbf{r},\mathbf{r}',\omega)$
$\!=-(4\pi|\mathbf{r}-\mathbf{r}'|)^{-1}$
$\!e^{-|\mathbf{r}-\mathbf{r}'|/\alpha(\omega)}$,
$\alpha(\omega)$ $\!=$ $\![\gamma(\omega)/
\sigma_\|(\omega)]^{1/2}$
$\!>$ $\!0$.
If the lack of exact knowledge of
$\tensor{K}(\mathbf{r},\mathbf{r}',\omega)$ really happens
to be an obstacle in an application, one can alternatively resort to the
approach on the basis of Eqs.~(\ref{eq306})--(\ref{eq308}),
by simply replacing therein
$\varepsilon(\omega)$ and $\kappa(\omega)$
by $\varepsilon(\mathbf{r},\omega)$ and
$\kappa(\mathbf{r},\omega)$, respectively.


\section{Green tensor construction for spatially dispersive bodies}
\label{sec5}

Practical application of the quantization scheme requires
the solution of the classical problem of
the determination of the Green tensor
$\tensor{G}(\mathbf{r},\mathbf{r}',\omega)$ for a given conductivity
tensor $\tensor{Q}(\mathbf{r},\mathbf{r}',\omega)$.
Typically, one has to deal with systems of
bodies each of which can be regarded
as being homogeneous in its respective interior region.
The physical surfaces of the bodies, including the boundary
surfaces between adjacent bodies, may be said to be
the particular space regions where the material properties
differ significantly from the
intra-body (bulk-material) properties.
Physical surfaces are therefore not mathematical ones but
more or less fuzzy boundary layers.
For not too small bodies, however, they usually
contain only a small fraction of the overall
material, so that they may often be approximately
replaced with (sharp) mathematical
surfaces, with the idealization that the
intra-body bulk-material
properties hold immediately beyond them.

\subsection{Dielectric approximation}
\label{sec5-1}

The application
of the point of view just outlined
to (systems of internally homogeneous) spatially dispersive bodies
is commonly referred to as the dielectric
approximation, for which it is obviously
required that the characteristic
length scale of spatial dispersion is small
in comparison with the typical
linear extensions of the bodies.
Hence, making use of the same notation as
employed in Eq.~(\ref{eq8-30}), we assume
the conductivity tensor of a system
of spatially dispersing bodies,
in the dielectric approximation,
to be of the form
\begin{equation}
\label{eq200}
\tensor{Q}(\mathbf{r,r'},\omega)
=
\theta[\mathbf{L(r')},\mathbf{r}]\,
\tensor{Q}[\mathbf{L(\mathbf{r'})},\mathbf{r-r'},\omega],
\end{equation}
with the vector $\mathbf{L}$
labeling now the bodies in place of the lattice cells in
Sec.~\ref{sec4A1}; $\mathbf{L}(\mathbf{r'})$
singles out the particular body that contains the
position $\mathbf{r'}$,
and the quantities
$\tensor{Q}[\mathbf{L},\mathbf{r-r'},\omega]$
are the bulk-material conductivity tensors
ascribed to the various bodies.
(Regions outside all actual bodies, if any, are formally
viewed as bodies in this notation.)
Note that
$\tensor{Q}(\mathbf{r,r'},\omega)$ as given by
Eq.~(\ref{eq200}) satisfies the reciprocity condition.
Corrections to the dielectric approximation
in the form of surface currents or, equivalently,
boundary conditions might be required for systems such as
needle-shaped bodies, thin films or the like, where the
dielectric approximation can be insufficient.
However, as such corrections
cannot be convincingly justified
without detailed (model) assumptions about the specific nature of the
physical surfaces, we do not consider them in the following.

\subsection{Integral equations and surface impedance method}
\label{sec5-2}
The dielectric approximation
renders it possible to formulate
integral equations from which the
Green tensor may be then derived.
Here we outline the surface impedance method,
which, in connection with its application
to the calculation of Casimir forces,
has recently given rise to
controversial discussions
(see, e.g.,
Refs.~\cite{GeyerB062003,EsquivelR062004,MiltonKA092004,MostepanenkoVM052006})
concerning the range of validity
and the approximations involved in this method.

To begin with, let us consider
spatially dispersive material described by a
conductivity tensor
$\tensor{Q}(\mathbf{r,r'},\omega)$
[not yet approximated by a form like Eq.~(\ref{eq200})],
and let $V$ denote some space region of interest.
Further, let
$\tensor{G}_\mathrm{aux}(\mathbf{r,r'},\omega)$ be the
Green tensor for an auxiliary problem to be specified yet.
Then, if $\fon{j}(\mathbf{r},\omega)$ is an arbitrarily
chosen current density
in the frequency domain, given inside
and/or
outside
$V$, and if $\fon{E}(\mathbf{r},\omega)$
and $\fon{B}(\mathbf{r,\omega})$ $\!=$
$\!(i\omega)^{-1}\Nabla\times\fon{E}(\mathbf{r},\omega)$,
respectively, are
the (classical) electric and induction fields
associated with this current density,
the identity
%
%
\begin{widetext}
\begin{multline}
\label{NL32}
\int_{V}\D^3r\,\fon{E}(\mathbf{r},\omega)\cdot
\Nabla\times\tensor{\Gamma}_\mathrm{aux}(\mathbf{r,r'},\omega)
+i\varepsilon_{0}\omega
\int_{V}\D^3r\,
\fon{E}(\mathbf{r},\omega)\cdot
\left[\tensor{G}_\mathrm{aux}(\mathbf{r,r'},\omega)
-(i\varepsilon_{0}\omega)^{-1}
\int_{V} \D^3s\,
\tensor{Q}(\mathbf{r,s},\omega)
\cdot\tensor{G}_\mathrm{aux}
(\mathbf{s,r'},\omega)\right]
\\
=\int_{V}\D^3r\,\fon{j}_{V}(\mathbf{r},\omega)\cdot
\tensor{G}_\mathrm{aux}(\mathbf{r,r'},\omega)+
%
%
\int_{\partial V} \D a(\mathbf{r})\,
[\fon{E}(\mathbf{r},\omega)\times\uv{n}(\mathbf{r})]\cdot
\tensor{\Gamma}_\mathrm{aux}(\mathbf{r,r',\omega})
\\
+
\mu_0^{-1}
\int_{\partial V}
\D a(\mathbf{r})\,
[\fon{
B
}
(\mathbf{r,\omega})\times
\uv{n}(\mathbf{r})]\cdot\tensor{G}_\mathrm{aux}(\mathbf{r,r'},\omega)
\end{multline}
\end{widetext}
(being a Green-type formula)
holds, as can be
proven correct by partial integration and
employing the reciprocity property of
$\tensor{Q}(\mathbf{r,s},\omega)$.
Here,
$da(\mathbf{r})$ and $\uv{n}(\mathbf{r})$ denote
the
absolute value
and the unit vector of the
surface element at $\mathbf{r}$ on the
surface $\partial V$ of $V$.
Further,
\begin{multline}
\label{NL34}
\fon{j}_{V}(\mathbf{r},\omega)
=
(i\mu_{0}\omega)^{-1}
\biggl[
\Nabla\times\Nabla\times\fon{E}(\mathbf{r},\omega)
\\
-
\frac{\omega^2}{c^2}\,\fon{E}(\mathbf{r},\omega)
-
i\mu_{0}\omega\int_{V}\D^3r'\,
\tensor{Q}(\mathbf{r,r'},\omega)
\cdot
\fon{E}(\mathbf{r}',\omega)
\biggr]
\end{multline}
and
\begin{equation}
\label{NL33}
\tensor{\Gamma}_\mathrm{aux}(\mathbf{r,r',\omega})
=(i
\mu_{0}
\omega)^{-1}
\Nabla\times\tensor{G}_\mathrm{aux}(\mathbf{r,r'},\omega).
\end{equation}
Note that $\fon{j}_{V}(\mathbf{r},\omega)$
does not agree with $\fon{j}(\mathbf{r},\omega)$,
in general, as the $\mathbf{r'}$-integral in Eq.~(\ref{NL34})
extends only over $V$.

Let us now assume that
$\tensor{Q}(\mathbf{r},\mathbf{r}',\omega)$
can be treated in the dielectric approximation
according to Eq.~(\ref{eq200}),
at least for the medium in $V$.
In this case, $\fon{j}_{V}(\mathbf{r},\omega)$
agrees with $\fon{j}(\mathbf{r},\omega)$ inside $V$,
and $\tensor{Q}(\mathbf{r},\mathbf{s},\omega)$ in
Eq.~(\ref{NL32}) can be replaced with the
bulk-medium conductivity tensor
$\tensor{Q}(\mathbf{L}_{V},\mathbf{r-s},\omega)$
attributed to the medium in $V$
($\mathbf{L}_{V}$ is the vector labeling region $V$).
Hence, if
$\tensor{G}_\mathrm{aux}(\mathbf{r,r',\omega})$
is required, for $\mathbf{r}\in V$ (at least),
to obey the equation
\begin{multline}
\label{eq300-a}
\Nabla\times\Nabla\times\tensor{G}_\mathrm{aux}(\mathbf{r,r'},\omega)
-
\frac{\omega^2}{c^2}\,\tensor{G}_\mathrm{aux}(\mathbf{r,r'},\omega)
\\
-
i\mu_{0}\omega\int_{V}\D^3s\,
\tensor{Q}(\mathbf{L}_{V},\mathbf{r-s},\omega)
\cdot
\tensor{G}_\mathrm{aux}(\mathbf{s,r'},\omega)
\\ =
\tensor{I}\delta(\mathbf{r-r'}),
\end{multline}
then Eq.~(\ref{NL32}) simplifies to
\begin{multline}
\label{NL37}
(i\mu_{0}\omega)^{-1}\fon{E}(\mathbf{r'},\omega)
\theta(\mathbf{L}_{V},\mathbf{r'})
=
(i\mu_{0}\omega)^{-1}\fon{E}^{\mathrm{(in)}}(\mathbf{r'},\omega)
\\
+
%
\int_{\partial V} \D a(\mathbf{r})\,
[\fon{E}(\mathbf{r},\omega)\times\uv{n}(\mathbf{r})]
\cdot
\tensor{\Gamma}_\mathrm{aux}(\mathbf{r,r'},\omega)
\\
+
\mu_0^{-1}
\int_{\partial V}\D a(\mathbf{r})\,
\bigl[\fon{
B
}
(\mathbf{r,\omega})\times\uv{n}(\mathbf{r})]
\cdot
\tensor{G}_\mathrm{aux}(\mathbf{r,r'},\omega),
\end{multline}
where
\begin{align}
\label{NL37-a}
\fon{E}^{\mathrm{(in)}}(\mathbf{r},\omega)
&
=
i\mu_{0}\omega \int_{V}\D^3r'\,
\fon{j}(\mathbf{r}',\omega)
\cdot
\tensor{G}_\mathrm{aux}(\mathbf{r',r},\omega)
\nonumber\\
&
=i\mu_{0}\omega\int_{V}\D^3r'\,
\tensor{G}_\mathrm{aux}(\mathbf{r,r'},\omega)
\cdot
\fon{j}(\mathbf{r}',\omega).
\end{align}
Since the characteristic length of spatial dispersion is assumed to be
sufficiently small as compared with the linear extensions of $V$,
we may extend, with little error, the $s$-integral in
Eq.~(\ref{eq300-a}) to the whole space and, therefore,
we may identify
$\tensor{G}_\mathrm{aux}(\mathbf{r,r'},\omega)$
in Eq.~(\ref{NL37})
with the corresponding bulk-medium Green tensor
as given in App.~\ref{AppD}.
Note that Eq.~(\ref{NL37}) may be viewed as
a statement of Huyghens' principle
(for $\mathbf{r'}$ inside $V$)
and of the extinction theorem (for $\mathbf{r'}$ outside $V$),
see Refs.~\cite{ChewBook,BornWolf},
which have been well known as a suitable starting point
for field calculations on the basis of
integral equation methods, see, in particular,
Refs.~\cite{AgarwalGS101972,AgarwalGS081974,AgarwalGS021975}
where the Wiener--Hopf technique
has been used to construct solutions
for a particular functional form
of $\tensor{Q}(\mathbf{L}_{V},\mathbf{r-s},\omega)$.

The method of surface impedance
(see, e.g., Ref.~\cite{MorseandFeshbach})
consists in the assumption that
a linear relation between the tangential field components of
$\fon{E}(\mathbf{r,\omega})$ and
$\fon{
B
}(\mathbf{r,\omega})$
exists on the surface $\partial V$,
\begin{multline}
\label{NL38a}
\uv{n}(\mathbf{r})\times\fon{E}(\mathbf{r},\omega)
\\
=
\mu_0^{-1}
\int_{\partial V} \D a(\mathbf{r'})
[\tensor{Z}(\mathbf{r,r',\omega})\times\uv{n}(\mathbf{r'})]
\cdot
[\fon{
B
}(\mathbf{r',\omega})
\times
\uv{n}(\mathbf{r'})]
%
\end{multline}
($\mathbf{r}$ on $\partial V$),
where the tensor $\tensor{Z}(\mathbf{r,r'},\omega)$ is the
dyadic surface impedance. Equivalently, one may consider the inverted form
\begin{multline}
\label{NL38b}
\mu_{0}^{-1}
\fon{
B
}(\mathbf{r,\omega})\times \uv{n}(\mathbf{r})
\\
=
\int_{\partial V} \D a(\mathbf{r'})
[\fon{E}(\mathbf{r'},\omega)\times \uv{n}(\mathbf{r'})]
\cdot
\tensor{Y}(\mathbf{r'},\mathbf{r},\omega)\cdot
\tensor{I}_{n}(\mathbf{r}),
%
\end{multline}
where $\tensor{I}_{n}(\mathbf{r})$
$\!=$ $\!\tensor{I}$ $\!-$
$\uv{n}(\mathbf{r})\uv{n}(\mathbf{r})$ is a
tangential projector, and
$\tensor{Y}(\mathbf{r,r',\omega})$
may be referred to as the surface admittance.
Once the relation (\ref{NL38a})
[or (\ref{NL38b})] has been adopted, one may convert
Eq.~(\ref{NL37}) (with $\mathbf{r'}\in V$)
into an integral equation
for $\fon{B}(\mathbf{r,\omega})$ [or $\fon{E}(\mathbf{r,\omega})$]
inside $V$, which can then be solved in terms of
$\fon{E}^{\mathrm{(in)}}(\mathbf{r},\omega)$
and the surface impedance, without specifying
the medium properties outside $V$.

At first glance, the method has some advantageous
features so that it has been enjoying
a reputation in the literature. One of these features
is that the necessity to know
the medium properties everywhere outside $V$ is
replaced, so to speak, by the necessity to
know the surface impedance for $\partial V$, which is
a great reduction at first glance.
Another one is that the application of continuity
conditions may be sidestepped to some extent.
Both of these points are to be qualified, however, and must be
seen in the context of the following remarks.

First, since the current density $\fon{j}(\mathbf{r},\omega)$
outside $V$ does
not contribute to Eq.~(\ref{NL37-a}),
the solution for $\fon{E}(\mathbf{r,\omega})$
can be unique only if
$\fon{j}(\mathbf{r},\omega)$ is located completely inside $V$;
solutions to the homogeneous integral equation can then be
excluded from consideration.
The surface impedance method applied to the volume $V$
can thus yield,
for given surface impedance,
the Green tensor $\tensor{G}(\mathbf{r,r'},\omega)$
only for both $\mathbf{r}$ and $\mathbf{r}'$
located in $V$
(which may be sufficient for
many applications, e.g., in the calculation
of dispersion forces).
Second, Eqs.~(\ref{NL38a}) and (\ref{NL38b})
implicitly demand that
the tangential components of the
electric and the induction
field uniquely exist on the surface $\partial V$ (i.e, they
should be continuous across the surface), or else ambiguities were
encountered. Quasi-local approximations of the
conductivity tensor should hence be
excluded if they contain magnetic-like singular
terms [as the second one in Eq.~(\ref{eq620})].
Provided that Eq.~(\ref{NL38a}) [or (\ref{NL38b})] holds,
the method requires knowledge of
$\tensor{Z}(\mathbf{r},\mathbf{r}',\omega)$
[or $\tensor{Y}(\mathbf{r},\mathbf{r}',\omega)$],
which plays the role of an external input.
If $\tensor{Z}(\mathbf{r},\mathbf{r}',\omega)$
[or $\tensor{Y}(\mathbf{r},\mathbf{r}',\omega)$]
is not known from the very beginning, one can
try to determine it \emph{a posteriori}
from the solution found, by appropriately specifying
the effect of the medium on the electromagnetic
outside the space region under consideration as well.
We proceed with an example where
this line can be pursued explicitly, on the basis of the
dielectric approximation.

\subsection{Example: Non-magnetic, planar systems}

Let $V$ denote
the slab-like region between two parallel planes
$z$ $\!=$ $\!0$ and $z$ $\!=$ $\!d$,
and let $\fon{j}(\mathbf{r,\omega})$ be located
completely inside $V$.
In order to apply the surface impedance method to the region
$V$, we assume that the medium in $V$ is non-magnetic
and can be treated in the dielectric approximation.
Introducing Fourier transforms according to
[$\mathbf{r}$ $\!\equiv$ $\!(\vec{\rho},z)$]
\begin{equation}
\label{NL41}
\fon{E}(\mathbf{r},\omega)=\int \D^2q\,
e^{i\mathbf{q}\cdot\vec{\rho}}
\fon{E}(z,\mathbf{q},\omega)
\end{equation}
%
%
and taking into account the lateral translational invariance
of the system, we may write Eq.~(\ref{NL38b}) in the Fourier
domain as
\begin{multline}
\label{NL40}
\mu_{0}^{-1}
\fon{
B}(z,\mathbf{q},\omega)\times\uv{z}
\\
=(2\pi)^2
\bigl\{
[\fon{E}(d,\mathbf{q},\omega)\times\uv{z}]\cdot
\tensor{Y}(d,z,-\mathbf{q},\omega)
\\
-[\fon{E}(0,\mathbf{q},\omega)\times
\uv{z}]\cdot
\tensor{Y}(0,z,-\mathbf{q},\omega)
\bigr\}\cdot\tensor{I}_{z}
\end{multline}
[$\tensor{I}_{z}$ $\!=$ $\!\tensor{I}$ $\!-$
$\!\uv{z}\uv{z}$],
where $z$ takes on the two values
$z$ $\!=$ $\!0$ and $z$ $\!=$ $\!d$
(the two terms enter with opposite
signs because of the opposite surface normals
on the two sides of $\partial V$).
Similarly, the Fourier transformed version of
Eq.~(\ref{NL37}) reads
\begin{multline}
\label{NL210}
\fon{E}(z',\mathbf{q},\omega)
\theta(\mathbf{L}_{V},z')
-\fon{E}^{\mathrm{(in)}}(z',\mathbf{q},\omega)
\\
=i\mu_{0}\omega (2\pi)^2
\bigl\{
[\fon{E}(d,\mathbf{q},\omega)\times\uv{z}]\cdot
\tensor{\Gamma}_\mathrm{aux}
(d,z',-\mathbf{q},\omega)
\\
-[\fon{E}(0,\mathbf{q},\omega)\times
\uv{z}]\cdot
\tensor{\Gamma}_\mathrm{aux}
(0,z',-\mathbf{q},\omega)
\\
+
\mu_{0}^{-1}
[\fon{
B}(d,\mathbf{q},\omega)\times\uv{z}]\cdot
\tensor{G}_\mathrm{aux}
(d,z',-\mathbf{q},\omega)
\\
-
\mu_{0}^{-1}
[\fon{
B
}(0,\mathbf{q},\omega)\times
\uv{z}]\cdot
\tensor{G}_\mathrm{aux}
(0,z',-\mathbf{q},\omega)
\bigr\}.
\end{multline}
Note that $\tensor{G}_\mathrm{aux}(\mathbf{r,r'},\omega)$
and
$\tensor{\Gamma}_{aux}(\mathbf{r,r'},\omega)$
are translationally invariant, because they
refer to bulk material. By
means of Eq.~(\ref{NL40}), Eq.~(\ref{NL210}) takes the form
\begin{multline}
\label{NL42}
\fon{E}(z',\mathbf{q},\omega)
\theta(\mathbf{L}_{V},z')
-\fon{E}^{\mathrm{(in)}}(z',\mathbf{q},\omega)
\\
=
\fon{E}(d,\mathbf{q},\omega)
\cdot
\tensor{I}_{z}
\cdot
\tensor{R}(d,z',\mathbf{q},\omega)
\\
-
\fon{E}(0,\mathbf{q},\omega)
\cdot
\tensor{I}_{z}
\cdot
\tensor{R}(0,z',\mathbf{q},\omega),
\end{multline}
with ($\zeta$ $\!=$ $\!0$, $\zeta$ $\!=$ $\!d$)
\begin{multline}
\label{NL43}
\tensor{R}(\zeta,z',\mathbf{q},\omega)
=i\mu_{0}\omega(2\pi)^2
\uv{z}\times
\Bigl\{
\tensor{\Gamma}_{1}(\zeta,z',-\mathbf{q},\omega)
\\
+(2\pi)^2 \tensor{Y}(\zeta,d,-\mathbf{q},\omega)\cdot
\tensor{I}_{z}
\cdot\tensor{G}_{1}(d,z',-\mathbf{q},\omega)
\\
-(2\pi)^2 \tensor{Y}(\zeta,0,-\mathbf{q},\omega)\cdot
\tensor{I}_{z}
\cdot\tensor{G}_{1}(0,z',-\mathbf{q},\omega)
\Bigr\}
\end{multline}
[note that
$\uv{z}\cdot\tensor{R}(\zeta,z',\mathbf{q},\omega)$ $\!=$ $\!0$].

From Eq.~(\ref{NL42}) it follows that (Appendix~\ref{AppE})
\begin{multline}
\label{NL44a}
\fon{E}(0,\mathbf{q},\omega)\cdot
\tensor{I}_{z}
=
[\fon{E}^{\mathrm{(in)}}(0+
,\mathbf{q},\omega)\cdot
\tensor{C}{^{\sharp}}
\\
-\fon{E}^{\mathrm{(in)}}(d-
,\mathbf{q},\omega
)\cdot\tensor{D}{^{\sharp}}
]
\cdot
[\tensor{A}\cdot\tensor{C}{^{\sharp}}
-\tensor{B}\cdot\tensor{D}{^{\sharp}}]{^{\sharp}},
\end{multline}
\begin{multline}
\label{NL44b}
\fon{E}(d,\mathbf{q},\omega)
\cdot
\tensor{I}_{z}
=
[
\fon{E}^{\mathrm{(in)}}(0+,\mathbf{q},\omega)
\cdot\tensor{A}{^{\sharp}}
\\
-
\fon{E}^{\mathrm{(in)}}(d-,\mathbf{q},\omega)
\cdot\tensor{B}{^{\sharp}}
]
\cdot
[\tensor{C}\cdot\tensor{A}{^{\sharp}}
-\tensor{D}\cdot\tensor{B}{^{\sharp}}]{^{\sharp}},
\end{multline}
where
\begin{align}
\label{NL50-a}
&\tensor{A}=\tensor{A}(\mathbf{q,\omega})=
\tensor{I}_{z}\cdot[
\tensor{I}+\tensor{R}(0,0+,\mathbf{q,\omega})]\cdot\tensor{I}_{z},
\\
\label{NL50-b}
&\tensor{B}=\tensor{B}(\mathbf{q,\omega})
=\tensor{I}_{z}\cdot\tensor{R}(0,d-,\mathbf{q,\omega})\cdot\tensor{I}_{z},
\\
\label{NL50-c}
&\tensor{C}=\tensor{C}(\mathbf{q,\omega})=
-\tensor{I}_{z}\cdot\tensor{R}(d,0+,\mathbf{q,\omega})
\cdot\tensor{I}_{z},
\\
\label{NL50-d}
&\tensor{D}=\tensor{D}(\mathbf{q,\omega})=
\tensor{I}_{z}\cdot[\tensor{I}-\tensor{R}(d,d-,\mathbf{q,\omega})]
\cdot\tensor{I}_{z},
\end{align}
and the superscript
$\sharp$ denotes matrix inversion
with respect to the $\tensor{I}_{z}$-space:
$\tensor{A}{^{\sharp}}$
$\!=$ $\!\tensor{I}_{z}\cdot(\tensor{I}_{z}\cdot\tensor{A}
\cdot\tensor{I}_{z})^{-1}\cdot\tensor{I}_{z}$.
The solution to Eq.~(\ref{NL42}) can now be readily
obtained by substituting
Eqs.~(\ref{NL44a}) and (\ref{NL44b}) back in
the right-hand side of Eq.~(\ref{NL42}).
Specifying in Eq.~(\ref{NL37-a}) the source of
$\fon{E}^\mathrm{(in)}(\mathbf{r},\omega)$
so as to correspond to a point dipole
situated in $V$,
$\fon{j}(\mathbf{r},\omega)$ $\!=$ $\!(i\mu_{0}\omega)^{-1}
\mathbf{p}\delta(\mathbf{r}$ $\!-$ $\!\mathbf{r}_{p})$,
from the corresponding position-space solution
$\fon{E}(\mathbf{r},\omega)
\theta(\mathbf{L}_{V},\mathbf{r})
$,
one can then read off, according to
$
\fon{E}(\mathbf{r},\omega)
\theta(\mathbf{L}_{V},\mathbf{r})$
$\!=$ $\!\mathbf{p}\cdot
\tensor{G}(\mathbf{r}_{p},\mathbf{r},\omega)
\theta(\mathbf{L}_{V},\mathbf{r})
\theta(\mathbf{L}_{V},\mathbf{r}_{p})
$,
the (interesting part of the) Green tensor
%
expressed in terms of the surface admittance,
i.e., in terms of
the quantities $\tensor{Y}(z,z',\mathbf{q},\omega)$
($z$ $\!=$ $\!0,d$%
%
;
$z'$ $\!=$ $\!0,d$).

In order to illustrate the calculation of
the admittance,
let us assume that the
(non-magnetic)
media to the left and right of $z$ $\!=$ $ \!0$ and
$z$ $\!=$ $ \!d$, respectively, are
(in the sense of
the dielectric approximation)
homogeneous half-spaces.
Using the indices $0$, $1$,
and $2$ to distinguish the left ($z\!<\!0$) half-space, the
volume $V$ ($0\!<\!z\!<\!d$), and the right half-space ($z\!>\!d$),
respectively, and associating with each of the three regions
the corresponding
bulk-medium Green tensor $\tensor{G}_{j}(\mathbf{r,r'},\omega)$
(App.~\ref{AppD}) and the associated auxiliary tensor
$\tensor{\Gamma}_{j}(\mathbf{r,r'},\omega)$
defined
according to
Eq.~(\ref{NL33}), one can derive the admittance by applying
the continuity conditions.
Introducing the abbreviations
\begin{align}
\label{NL51-a}
&
\tensor{M}(\mathbf{q,\omega})=
-
\tensor{I}_{z}\cdot
\tensor{G}_{1}(0,d-,\mathbf{q,\omega})
\cdot\tensor{I}_{z},
\\
\label{NL51-b}
&
\tensor{N}(\mathbf{q,\omega})
=-\tensor{I}_{z}\cdot
[\tensor{G}_{1}(0,0+,\mathbf{q,\omega})
+\tensor{G}_{0}(0,0-,\mathbf{q,\omega})
]
\cdot\tensor{I}_{z},
\end{align}
\begin{align}
\label{NL51-c}
&
\tensor{P}(\mathbf{q,\omega})=
\tensor{I}_{z}\cdot
[\tensor{G}_{1}(d,d-,\mathbf{q,\omega})
+\tensor{G}_{2}(0,0+,\mathbf{q,\omega})
]
\cdot\tensor{I}_{z},
\\
\label{NL51-d}
&
\tensor{S}(\mathbf{q,\omega})=
\tensor{I}_{z}\cdot
\tensor{G}_{1}(d,0+,\mathbf{q,\omega})
\cdot\tensor{I}_{z},
\\
\label{NL52-a}
&
\tensor{T}(\mathbf{q,\omega})=
\tensor{I}_{z}\cdot
\tensor{\Gamma}_{1}(0,d-,\mathbf{q,\omega})
\cdot\tensor{I}_{z},
\end{align}
\begin{align}
\label{NL52-b}
&
\tensor{U}(\mathbf{q,\omega})
=\tensor{I}_{z}\cdot
[\tensor{\Gamma}_{1}(0,0+,\mathbf{q,\omega})
+\tensor{\Gamma}_{0}(0,0-,\mathbf{q,\omega})
]
\cdot\tensor{I}_{z},
\\
\label{NL52-c}
&
\tensor{V}(\mathbf{q,\omega})=
-
\tensor{I}_{z}\cdot
[\tensor{\Gamma}_{1}(d,d-,\mathbf{q,\omega})
+\tensor{\Gamma}_{2}(0,0+,\mathbf{q,\omega})
]
\cdot\tensor{I}_{z},
\\
\label{NL52-d}
&
\tensor{W}(\mathbf{q,\omega})=
-
\tensor{I}_{z}\cdot
\tensor{\Gamma}_{1}(d,0+,\mathbf{q,\omega})
\cdot\tensor{I}_{z},
\end{align}
one eventually finds (App.~\ref{AppF})
the $\tensor{Y}(z,z',\mathbf{q},\omega)$
($z$ $\!=$ $\!0,d$%
;
$z'$ $\!=$ $\!0,d$) to be
(for notational convenience, the arguments
$\mathbf{q}$, $\omega$ are suppressed)
%
\begin{align}
\label{NL53-a}
&\tensor{Y}(0,0)=
\frac{1}{(2\pi)^2}
[\tensor{U}\cdot\tensor{S}{^{\sharp}}
-\tensor{T}\cdot\tensor{P}{^{\sharp}}]
\cdot
[\tensor{M}\cdot\tensor{P}{^{\sharp}}
-\tensor{N}\cdot\tensor{S}{^{\sharp}}
]{^{\sharp}},
\\
\label{NL53-b}
&\tensor{Y}(0,d)=
\frac{1}{(2\pi)^2}
[\tensor{U}\cdot\tensor{N}{^{\sharp}}
-\tensor{T}\cdot\tensor{M}{^{\sharp}}]
\cdot
[\tensor{P}\cdot\tensor{M}{^{\sharp}}
-\tensor{S}\cdot\tensor{N}{^{\sharp}}
]{^{\sharp}},
\\
\label{NL53-c}
&\tensor{Y}(d,0)=\frac{1}{(2\pi)^2}
[\tensor{V}\cdot\tensor{P}{^{\sharp}}
-\tensor{W}\cdot\tensor{S}{^{\sharp}}]
\cdot
[\tensor{M}\cdot\tensor{P}{^{\sharp}}
-\tensor{N}\cdot\tensor{S}{^{\sharp}}
]{^{\sharp}},
\\
\label{NL53-d}
&\tensor{Y}(d,d)=\frac{1}{(2\pi)^2}
[\tensor{V}\cdot\tensor{M}{^{\sharp}}
-\tensor{W}\cdot\tensor{N}{^{\sharp}}]
\cdot
[\tensor{P}\cdot\tensor{M}{^{\sharp}}
-\tensor{S}\cdot\tensor{N}{^{\sharp}}
]{^{\sharp}},
\end{align}
which are, quite naturally, far more complicated
than the simple Kliever--Fuchs
expressions mentioned in App.~\ref{AppD}.
%

\section{Summary and concluding remarks}
\label{sec6}
We have developed a rather general quantization scheme for the
macroscopic electromagnetic field in arbitrary linearly responding
media, which offers a unified approach to QED in linear media.
Describing the medium response by a non-local conductivity tensor,
any of the possible
electromagnetic features of a linear medium in equilibrium
is covered by the scheme, in particular, spatial dispersion.
Central quantities of the scheme are the noise current that
is intimately connected with the absorption necessarily observed in
any linear medium in equilibrium, the bosonic dynamical variables
associated with the noise current, and
the Green tensor of the phenomenological Maxwell
equations, in which the medium properties enter via
the conductivity tensor.

From a careful analysis of the
dynamical variables and (quasi-)local limiting forms of
the non-local conductivity tensor, we have shown how 
quantization schemes  previously developed for locally responding
media can be recovered as special applications of the general
quantization scheme.
In particular,
a locally responding magnetodielectric medium can be 
viewed as a special quasi-local limiting case
of an isotropic, spatially dispersive medium without optical 
activity, where the (local) dielectric permittivity and 
magnetic permeability are just
two contributions to one and the same quasi-local 
conductivity tensor.
As a result, 
application of the general quantization scheme shows that
the electromagnetic field in such a medium 
can be quantized by using a single set of bosonic variables.

Generally, the use of a single set of bosonic variables means that the 
noise current which enters
the macroscopic Maxwell equations
is not divided into parts
(associated, e.g., with a polarization and a magnetization)
regarded as representing independent
degrees of freedom,
but is rather treated as an entity.
This may be particularly advantageous for future studies of (quantum)
electrodynamics in moving media, simplifying 
the discussion of transformations to different frames of reference.
However, the theory also admits, by appropriate projection, 
the use of several independent sets
of bosonic variables, 
which in fact corresponds to the neglect of certain kinds of
interactions in the sense of super-selection rules. 

Since exact solutions of Maxwell's equations are
not available in closed form in general,
even more so if spatial dispersion is taken into account, one has
to resort to approximation methods to obtain explicit expressions 
for the Green tensor, the latter being one of the cornerstones of the
theory. To assist in such intentions we have considered in some
detail the dielectric approximation for the
conductivity tensor, which consists in approximating
the conductivity tensor of a system of spatially dispersive bodies by
joining together
bulk-medium conductivity tensors (which are
routinely handled in reciprocal space).
Although the information relevant to
physical surface regions is lost
in this way, the dielectric approximation has been 
a key tool to render tractable
electromagnetic propagation problems in spatially dispersive media.
Accepting the dielectric approximation, the problem of finding the
Green tensor becomes then solvable
via integral-equation and surface-impedance techniques.

As already mentioned, diamagnetic media are not covered
by the quantization scheme developed in this paper---a scheme 
that exhausts the possibilities offered by the linear-response 
framework.
Furthermore, it should be pointed out that the scheme
does also not automatically apply to linearly amplifying
media. Although both types of media 
do not really fit into the linear-response framework,
they may be forced into it, 
but not without reservations and alterations of the whole scheme.
Clearly, the concept of linear amplification has
a range of applicability very much smaller than that of
linear dissipation. 

Concluding, this work provides the most general quantization scheme
for the electromagnetic field in linearly responding, absorbing
materials to date, from which
previously given schemes can be
recovered as limiting cases.
It will serve as a foundation for
investigations of surface plasmon effects involving strong spatial
dispersion, and as a starting point for the investigation of moving media.


\begin{appendix}
\section{Derivation of Eq.~(\ref{eq8-3})}
\label{AppA}

The linear integro-differential equation
(\ref{eq8-3}) can be represented as
\begin{equation}
\label{NL8}
\int \D^3s\,
\tensor{H}(\mathbf{r},\mathbf{s},\omega)
\cdot
\tensor{G}(\mathbf{s},\mathbf{r}',\omega)
=
\tensor{I}\delta(\mathbf{r}-\mathbf{r}'),
\end{equation}
where the integral kernel
\begin{multline}
\label{NLA0}
\tensor{H}(\mathbf{r},\mathbf{r}',\omega)
=
\Nabla\times\Nabla\times\tensor{I}\delta(\mathbf{r}-\mathbf{r}')
\\
-\frac{\omega^2}{c^2}\tensor{I}
\delta(\mathbf{r}-\mathbf{r}')
-i\mu_{0}\omega\,\tensor{Q}(\mathbf{r},\mathbf{r}',\omega)
\end{multline}
is reciprocal,
\begin{equation}
\label{NLA10}
\tensor{H}(\mathbf{r},\mathbf{r}',\omega)
=\tensor{H}{^\mathsf{T}}(\mathbf{r}',\mathbf{r},\omega),
\end{equation}
since $\tensor{Q}(\mathbf{r},\mathbf{r}',\omega)$ is reciprocal.
Hence, the transposed equation of
Eq.~(\ref{NL8}) takes the form
\begin{equation}
\label{NLA1}
\int \D^3s\,
\tensor{G}{^\mathsf{T}}(\mathbf{s},\mathbf{r},\omega)
\cdot
\tensor{H}(\mathbf{s},\mathbf{r'},\omega)
=
\tensor{I}
\delta(\mathbf{r}-\mathbf{r}').
\end{equation}
Multiplying from the right with
$\tensor{G}(\mathbf{r}',\mathbf{s'},\omega)$, integrating over
$\mathbf{r'}$, and using Eq.~(\ref{NL8}),
one can see that the Green tensor is also reciprocal,
\begin{equation}
\label{NLA2}
\tensor{G}(\mathbf{r},\mathbf{r}',\omega)
=\tensor{G}{^\mathsf{T}}(\mathbf{r}',\mathbf{r},\omega).
\end{equation}
Because of Eq.~(\ref{NLA2}), the complex conjugate of Eq.~(\ref{NLA1})
reads
\begin{equation}
\label{NLA3}
\int \D^3s\,
\tensor{G}{^\ast}(\mathbf{r},\mathbf{s},\omega)
\cdot
\tensor{H}{^\ast}(\mathbf{s},\mathbf{r}',\omega)
=\tensor{I}\delta(\mathbf{r}-\mathbf{r}').
\end{equation}
Taking the dot product of Eq.~(\ref{NL8}) from the left with
$\tensor{G}{^{\ast}}(\mathbf{s}',\mathbf{r},\omega)$ and
integrating over $\mathbf{r}$,
taking the dot product of Eq.~(\ref{NLA3}) from the right with
$\tensor{G}(\mathbf{r}',\mathbf{s}')$ and
integrating over $\mathbf{r}'$, and subtracting
the two resulting equations, one derives
\begin{multline}
\label{NLA5}
\Im\tensor{G}(\mathbf{r},\mathbf{r}',\omega)=
\\
-
\int \D^3s\int \D^3s' \,\tensor{G}(\mathbf{r},\mathbf{s},\omega)
\cdot[\Im \tensor{H}(\mathbf{s},\mathbf{s}',\omega)]
\cdot\tensor{G}{^\ast}(\mathbf{s}',\mathbf{r}',\omega).
\end{multline}
From Eq.~(\ref{NLA0}) it is seen that
\begin{multline}
\label{NLA6}
\Im \tensor{H}(\mathbf{r},\mathbf{r}',\omega)
=
\\
-\frac
{\Im \omega^2}{c^2}\,
\tensor{I}\delta(\mathbf{r}-\mathbf{r}')
-
\mu_{0}\Re[\omega\tensor{Q}(\mathbf{r},\mathbf{r}',\omega)].
\end{multline}
Insertion of
Eq.~(\ref{NLA6}) in Eq.~(\ref{NLA5})
and restriction to real frequencies
leads, upon recalling Eq.~(\ref{eq2}),
to Eq.~(\ref{eq8-3}).


\section{Non-uniqueness of the kernel
${\protect\tensor{\bm{K}}\bm{(\mathbf{r,r'},\omega)}}$}
\label{AppB}

The transition from $\tensor{K}(\mathbf{r},\mathbf{r}',\omega)$ to
$\tensor{K}{'}(\mathbf{r},\mathbf{r}',\omega)$ according to
Eq.~(\ref{NL21-1}) can be re-interpreted as a redefinition
of the dynamical variables
$\hat{\mathbf{f}}(\mathbf{r},\omega)$
and $\hat{\mathbf{f}}^\dagger(\mathbf{r},\omega)$ according to
\begin{align}
\label{B1-1}
&
\hat{\mathbf{f}}(\mathbf{r},\omega)
= \int\D^3r'\,\tensor{V}(\mathbf{r},\mathbf{r}',\omega)
\cdot\hat{\mathbf{f}}'(\mathbf{r}',\omega),
\\
\label{B1-2}
&
\hat{\mathbf{f}}^\dagger(\mathbf{r},\omega)
=
\int\D^3r'\,\tensor{V}{^\ast}(\mathbf{r},\mathbf{r}',\omega)
\cdot\hat{\mathbf{f}}^{\prime\dagger}(\mathbf{r}',\omega).
\end{align}
Inserting Eq.~(\ref{B1-1}) into Eq.~(\ref{eq8-5}) yields
\begin{equation}
\label{B2}
\fo{j}_{\mathrm N}(\mathbf{r},\omega)
=
\left(\frac{\hbar\omega}{\pi}\right)^{\frac{1}{2}}
\int \D^3r'\, \tensor{K}{'}(\mathbf{r},\mathbf{r}',\omega)
\cdot
\hat{\mathbf{f}}'(\mathbf{r}',\omega),
\end{equation}
where $\tensor{K}{'}(\mathbf{r},\mathbf{r}',\omega)$
is just given by Eq.~(\ref{NL21-1}).
With regard to the transformation (\ref{B1-1}) and (\ref{B1-2}), the
significance of replacing
Eq.~(\ref{NL21-2}) with Eq.~(\ref{B3})
is that the variables
$\hat{\mathbf{f}}'(\mathbf{r},\omega)$ and
$\hat{\mathbf{f}}^{\prime\dagger}(\mathbf{r},\omega)$
are uniquely expressible in terms of the
$\hat{\mathbf{f}}(\mathbf{r},\omega)$ and
$\hat{\mathbf{f}}^\dagger(\mathbf{r},\omega)$, and so are
on an equal footing with them---%
the unitary operator associated with
$\tensor{V}(\mathbf{r},\mathbf{r}',\omega)$
uniquely maps a set of bosonic variables onto
a fully equivalent set of bosonic variables.
Hence, $\tensor{V}(\mathbf{r},\mathbf{r}',\omega)$ may
be thought of as being included in the chosen set of dynamical
variables. In this sense, it is sufficient to base the
calculations in Sec.~\ref{sec3} on the Hermitian
operator associated with the integral kernel
$\tensor{K}(\mathbf{r},\mathbf{r}',\omega)$
as defined by Eq.~(\ref{NL21}).

It is worth noting that the operator
associated with
$\tensor{K}{'}(\mathbf{r},\mathbf{r},\omega)$ as
defined by Eq.~(\ref{NL21-1}) is non-Hermitian
whenever $\tensor{V}(\mathbf{r},\mathbf{r}',\omega)$ is non-trivial.
To see this, let us conversely assume that the
operator
associated with
$\tensor{K}{'}(\mathbf{r},\mathbf{r},\omega)$ is Hermitian,
\begin{multline}
\label{B7}
\int \D^3s\,
\tensor{K}(\mathbf{r,s},\omega)
\cdot
\tensor{V}(\mathbf{s,r'},\omega)
\\
=
\int \D^3s\,
\tensor{V}{^{+}}(\mathbf{s,r},\omega)
\cdot
\tensor{K}(\mathbf{s,r'},\omega).
\end{multline}
Applying from the left the
operator
associated with $\tensor{V}$ and from the right the
operator associated with $\tensor{V}{^{+}}$
and recalling Eq.~(\ref{NL21-2}), one sees that
\begin{multline}
\label{B8}
\int \D^3s\,
\tensor{V}(\mathbf{r,s},\omega)
\cdot
\tensor{K}(\mathbf{s,r'},\omega)
\\
=
\int \D^3s\,
\tensor{K}(\mathbf{r,s},\omega)
\cdot
\tensor{V}{^{+}}(\mathbf{r',s},\omega).
\end{multline}
Applying the
operator associated with $\tensor{K}$ from the left to Eq.~(\ref{B7})
and from the right to Eq.~(\ref{B8}) and comparing the results,
one finds that the
operators
associated with
$\tensor{V}
$ and
$\tensor{\sigma}
$
commute [recall Eq.~(\ref{eq8-7})],
so that the
operator
associated with $\tensor{V}$
maps each (possibly degenerate) eigenspace of the
operator
associated with
$\tensor{\sigma}
$
onto itself. Specifically, this implies that
the operator
associated with
$\tensor{V}
$
commutes with the spectral projectors of the
operator
associated with
$
\tensor{\sigma}
$
[and, therefore, also
with the projectors (\ref{NL18-1})].
Since the spectral projectors of the
operators
associated with
$
\tensor{\sigma}
$
and
$\tensor{K}
$
are the same
[cf.~Eqs.~(\ref{NL20}) and (\ref{NL21})], the
operators
associated with
$\tensor{V}
$
and
$\tensor{K}
$
also commute.
But then, since
the operator
associated with
$\tensor{K}
$
is
invertible (being a positive operator),
Eq.~(\ref{B7}) [or Eq.~(\ref{B8})]
shows that the
operator
associated with
$\tensor{V}
$ is Hermitian,
i.e.,
\begin{equation}
\label{B10}
\tensor{V}(\mathbf{r},\mathbf{r}',\omega)=
\tensor{V}{^{+}}(\mathbf{r}',\mathbf{r},\omega).
\end{equation}
Since the operator
associated with $\tensor{V}$ is also unitary,
in the diagonal expansion
\begin{equation}
\label{B4}
\tensor{V}(\mathbf{r},\mathbf{r}',\omega)
=\int \D\alpha\,
v(\alpha,\omega)\,
\mathbf{F}(\alpha,\mathbf{r},\omega)
\ot
\mathbf{F}^{\ast}(\alpha,\mathbf{r}',\omega),
\end{equation}
one must have $v(\alpha,\omega)$ $\!=$ $\!\pm 1$
for each $\alpha$,
which means that the
Hermitian operator associated with
$\tensor{K}{'}(\mathbf{r},\mathbf{r}',\omega)$
can differ from
the operator associated with
$\tensor{K}(\mathbf{r},\mathbf{r}',\omega)$ only by
the trivial type of unitary transformation
that merely replaces some of the
basis functions
$\mathbf{F}(\alpha,\mathbf{r},\omega)$
with
$-\mathbf{F}(\alpha,\mathbf{r},\omega)$.
Conversely, this shows that any
(in this sense) non-trivial
$\tensor{V}(\mathbf{r},\mathbf{r}',\omega)$ necessarily yields a
non-Hermitian $\tensor{K}{'}(\mathbf{r},\mathbf{r}',\omega)$.

\section{Reduced state space
and super-selection rule}
\label{AppC}
Let us consider the state space spanned by the
Fock states associated with
$\hat{\mathbf{f}}(\mathbf{r},\omega)$ and
$\hat{\mathbf{f}}^{\dagger}(\mathbf{r},\omega)$
so that an arbitrary, normalizable state
$|\phi\rangle$
in this space can be represented
in the form
\begin{multline}
\label{C1}
\ket{\phi}
=
\ket{0}\!\EW{0|\phi}
\\
+\sum_{k_{1}=1}^{3}\int_{0}^{\infty}\D\omega_{1}\int\D^3r_{1}\,
\phi_{k_1}(\mathbf{r}_{1},\omega_{1})
\ket{1_{k_1}(\mathbf{r}_{1},\omega_{1})}
%
\\
+
\sum_{k_1,k_2=1}^{3}
\int_{0}^{\infty}\D\omega_{1}
\int_{0}^{\infty}\D\omega_{2}\int\D^3r_{1}
\int\D^3r_{2}\,
\\
\times
\phi_{k_1 k_2}(\mathbf{r}_{1},\omega_{1},\mathbf{r}_{2},\omega_{2})
\ket{1_{k_{1}}(\mathbf{r}_{1},\omega_{1}),1_{k_{2}}(\mathbf{r}_{2},\omega_{2})}
\\
+\ldots,
%
\end{multline}
where
\begin{align}
\label{C2}
&
\hat{f}_{k}(\mathbf{r},\omega)\ket{0}=0,
\\
\label{C3}
&
\hat{f}_{k}^{\dagger}(\mathbf{r},\omega)\ket{0}
=\ket{1_{k}(\mathbf{r},\omega)},
\\
\label{C4}
&
\hat{f}_{k_{N}}^{\dagger}(\mathbf{r}_{N},\omega_{N})\cdots
\hat{f}_{k_{1}}^{\dagger}(\mathbf{r}_{1},\omega_{1})\ket{0}
\nonumber\\
&\hspace{3ex}
=\ket{1_{k_{1}}(\mathbf{r}_{1},\omega_{1}),
\dots,1_{k_{N}}(\mathbf{r}_{N},\omega_{N})}.
\end{align}
The normalization of $|\phi\rangle$ can be obtained
by using the formula (which can be viewed as a
special case of the Bloch--De~Dominicis
theorem \cite{KuboNonEqStat})
\begin{multline}
\bra{0}
\hat{f}_{k_{M}}(\mathbf{r}_{M},\omega_{M})
\ldots
\hat{f}_{k_{1}}(\mathbf{r}_{1},\omega_{1})
\\\times
\hat{f}_{k_{1}'}^{\dagger}(\mathbf{r}_{1}',\omega_{1}')
\ldots
\hat{f}_{k_{N}'}^{\dagger}(\mathbf{r}_{N}',\omega_{N}')
\ket{0}
\\
=
\delta_{MN}\sum_{\pi \in \mathcal{S}_N}
\prod_{l=1}^{N}\delta_{k_{l},k_{\pi(l)}'}
\delta\bigl(\mathbf{r}_{l}-\mathbf{r}_{\pi(l)}'\bigr)
\delta\bigl(\omega_{l}-\omega_{\pi(l)}'\bigr)
\end{multline}
($\langle 0|0\rangle$ $\!=$ $\!1$;
$\mathcal{S}_{N}$, group of
permutations of $N$ objects).

In order to construct a reduced state
space in which the operators
$\hat{\mathbf{f}}_{\lambda}(\mathbf{r},\omega)$ and
$\hat{\mathbf{f}}_{\lambda}^{\dagger}(\mathbf{r'},\omega')$
defined by Eq.~(\ref{NL25d}) behave like bosonic
operators, let us first introduce states $\ket{0}_{\lambda}$
according to
\begin{align}
\label{C5-1}
&
\hat{f}_{\lambda i}(\mathbf{r},\omega)\ket{0}_{\lambda}
=0,
\\
\label{C5-3}
&
\hat{f}_{\lambda i}(\mathbf{r},\omega)\ket{0}_{\lambda'}
=
\ket{0}_{\lambda'}\!\hat{f}_{\lambda i}(\mathbf{r},\omega)\qquad
(\lambda \ne \lambda')
\end{align}
($_\lambda\langle 0|0\rangle_\lambda$ $\!=$ $\!1$), such that
\begin{equation}
\label{C5-2}
\ket{0}
=
\bigotimes_{\lambda=1}^{\Lambda}\ket{0}_{\lambda}.
\end{equation}
Now let us introduce, for each $\lambda$, an orthogonal
projector $\hat{P}_{\lambda}$ as the
sum of orthogonal projectors
$\hat{P}_{\lambda}^{(N)}$,
\begin{align}
\label{C8}
&
\hat{P}_{\lambda}=\sum_{N=0}^{\infty}\hat{P}_{\lambda}^{(N)},
\\
\label{C9}
&\hat{P}_{\lambda}^{(N)\dagger}=\hat{P}_{\lambda}^{(N)},
\\
\label{C10}
&
\hat{P}_{\lambda}^{(N)}\hat{P}_{\lambda}^{(N')}
=\delta_{NN'}\hat{P}_{\lambda}^{(N)}
\end{align}
and specify $\hat{P}_{\lambda}^{(N)}$ in such a way
that, when applied to a quantum state of the form (\ref{C1}),
it picks out the \mbox{$(N$ $\!+$ $\!1)$th} term on
the right-hand side of Eq.~(\ref{C1}) and incorporates $N$
position-space projection kernels belonging to the
chosen value of $\lambda$,
\begin{equation}
\label{C11}
\hat{P}_{\lambda}^{(0)}
=
\ket{0}_{\lambda}
\!
\bra{0}_{\lambda}
\end{equation}
\begin{widetext}
\begin{multline}
\label{C14}
\hat{P}_{\lambda}^{(N)}
=\frac{1}{N!}
\sum_{k_{1}}\int_{0}^{\infty}\D\omega_{1}\int \D^3r_{1}
\sum_{k_{2}}\int_{0}^{\infty}\D\omega_{2}\int \D^3r_{2}\cdots
\sum_{k_{N}}\int_{0}^{\infty}\D\omega_{N}\int \D^3r_{N}
\\
\times
\hat{f}_{\lambda k_{1}}^{\dagger}(\mathbf{r}_{1},\omega_{1})
\hat{f}_{\lambda k_{2}}^{\dagger}(\mathbf{r}_{2},\omega_{2})\cdots
\hat{f}_{\lambda k_{N}}^{\dagger}(\mathbf{r}_{N},\omega_{N})
\hat{P}_{\lambda}^{(0)}
\hat{f}_{\lambda k_{N}}(\mathbf{r}_{N},\omega_{N})
\hat{f}_{\lambda k_{N-1}}(\mathbf{r}_{N-1},\omega_{N-1})
\cdots
\hat{f}_{\lambda k_{1}}(\mathbf{r}_{1},\omega_{1})
\end{multline}
\end{widetext}
($N\!=\!1,2,\ldots$). It is not difficult to prove
that Eqs.~(\ref{C9}) and (\ref{C10})
are fulfilled, where the latter equation fixes the
normalization factor $1/N!$ in Eq.~(\ref{C14}), and that,
in view of Eqs.~(\ref{C5-2}) and (\ref{C10}),
the commutation relation
\begin{equation}
\label{C14-1}
\bigl[\hat{P}_{\lambda}^{(N)},\hat{P}_{\lambda'}^{(N')}\bigr]=0
\end{equation}
holds.
We may now define a reduced state space that
contains only those (normalizable) vectors that have
the separable form
\begin{align}
\label{C22}
&
\ket{\phi}^\mathrm{(red)}
=\bigotimes_{\lambda=1}^{\Lambda}\ket{\phi}_{\lambda},
\\
\label{C23}
&
\hat{P}_{\lambda} \ket{\phi}_{\lambda}
=\ket{\phi}_{\lambda},
\end{align}
with each vector
$\ket{\phi}_{\lambda}$ being, by construction,
a superposition of vectors
\begin{multline}
\label{C30}
\ket{N}_{\lambda}
=
\sum_{k_{1}}\int_{0}^{\infty}\D\omega_{1}\int \D^3r_{1}
\ldots
\sum_{k_{N}}\int_{0}^{\infty}\D\omega_{N}\int \D^3r_{N}
\\ \times
C_{\lambda k_{1}\ldots\lambda k_{N}}
(\mathbf{r}_{1},\omega_{1},\cdots,\mathbf{r}_{N},\omega_{N})
\\ \times
\ket{1_{\lambda k_{1}}(\mathbf{r}_{1},\omega_{1}),\ldots,
1_{\lambda k_{N}}(\mathbf{r}_{N},\omega_{N})},
\end{multline}
where, in analogy to Eq.~(\ref{C4}),
\begin{multline}
\label{C30-1}
\ket{1_{\lambda k_{1}}(\mathbf{r}_{1},\omega_{1}),
\dots,1_{\lambda k_{N}}(\mathbf{r}_{N},\omega_{N})}
\\
=
\hat{f}_{\lambda k_{N}}^{\dagger}(\mathbf{r}_{N},\omega_{N})\cdots
\hat{f}_{\lambda k_{1}}^{\dagger}(\mathbf{r}_{1},\omega_{1})
\ket{0}_{\lambda}
.
\end{multline}
The important feature of these states is that
the result of performing the integrations in Eq.~(\ref{C30})
is not changed if the wave function
$C_{\lambda k_{1}\ldots\lambda k_{N}}
(\mathbf{r}_{1},\omega_{1},\cdots,\mathbf{r}_{N},\omega_{N})$
is replaced according to
\begin{multline}
\label{C19-1}
C_{\lambda k_{1}\ldots\lambda k_{N}}
(\mathbf{r}_{1},\omega_{1},\cdots,\mathbf{r}_{N},\omega_{N})
\\
\mapsto
\int\D^3r_{1}'\cdots\int\D^3r_{N}'
\,
(\tensor{P}_{\lambda})_{k_{1}k_{1}'}(\mathbf{r}_1,\mathbf{r}_1',\omega_1)\cdots
\\ \times
(\tensor{P}_{\lambda})_{k_{N}k_{N}'}(\mathbf{r}_N,\mathbf{r}_N',\omega_N)
C_{\lambda k_{1}'\ldots\lambda k_{N}'}
(\mathbf{r}_{1}',\omega_{1},\cdots\!,\mathbf{r}_{N}',\omega_{N}).
\end{multline}
It is also not changed if
$C_{\lambda k_{1}\ldots\lambda k_{N}}
(\mathbf{r}_{1},\omega_{1},\cdots,\mathbf{r}_{N},\omega_{N})$
is symmetrized with repect to the labels $1,\ldots,N$.
Wave functions that can be reduced
to the same standardized wave function by these operations
are thus fully equivalent representatives of the same vector.
Without loss of generality, one can thus adopt the
convention to employ only
such standardized wave functions.

The commutation relation (\ref{eq8-16}) implies that
\begin{multline}
\label{C40}
e^{\epsilon \hat{f}_{\lambda k}(\mathbf{r},\omega)}
\hat{f}_{\lambda' k'}^{\dagger}(\mathbf{r}',\omega')
e^{- \epsilon \hat{f}_{\lambda k}(\mathbf{r},\omega)}
\\
=\hat{f}_{\lambda' k'}^{\dagger}(\mathbf{r}',\omega')
+
\epsilon \delta_{\lambda\lambda'}
(\tensor{P}_{\lambda})_{k k'}
(\mathbf{r},\mathbf{r}',\omega)\delta(\omega-\omega')
\end{multline}
with $\epsilon$ being a parameter. As Eq.~(\ref{C40}) is a similarity
transformation, it generalizes to
\begin{multline}
\label{C41}
e^{\epsilon \hat{f}_{\lambda k}(\mathbf{r},\omega)}
F[\hat{f}_{\lambda' k'}^{\dagger}(\mathbf{r}',\omega')]
e^{-\epsilon \hat{f}_{\lambda k}(\mathbf{r},\omega)}
\\
=
F[\hat{f}_{\lambda' k'}^{\dagger}(\mathbf{r}',\omega')
+
\epsilon \delta_{\lambda\lambda'}
(\tensor{P}_{\lambda})_{k k'}
(\mathbf{r},\mathbf{r}',\omega)
\delta(\omega-\omega')]
\end{multline}
where
$F=F[\hat{f}_{\lambda' k'}^{\dagger}(\mathbf{r}',\omega')]$ is
any well-behaved functional of
$\hat{f}_{\lambda' k'}^{\dagger}(\mathbf{r}',\omega')$.
Comparison of the terms of first order in $\epsilon$ on both sides
yields
\begin{multline}
\label{C42}
\left[
\hat{f}_{\lambda k}(\mathbf{r},\omega),
F[\hat{f}_{\lambda' k'}^{\dagger}(\mathbf{r}',\omega')]
\right]
\\
=
\bigg\{
\frac{\partial}{\partial\epsilon}
F[\hat{f}_{\lambda' k'}^{\dagger}(\mathbf{r}',\omega')
\\+
\epsilon \delta_{\lambda\lambda'}
(\tensor{P}_{\lambda})_{k k'}
(\mathbf{r},\mathbf{r}',\omega)
\delta(\omega-\omega')]
\bigg\}_{\epsilon=0}.
\end{multline}
Let us consider the particular functional
$F_{N}[\hat{f}_{\lambda' k'}^{\dagger}(\mathbf{r}',\omega')]$
appearing in Eqs.~(\ref{C30}), (\ref{C30-1}),
\begin{multline}
\label{C43}
F_{N}[\hat{f}_{\lambda' k'}^{\dagger}(\mathbf{r}',\omega')]
\\
=
\sum_{k_{1}}\int_{0}^{\infty}\D\omega_{1}\int \D^3r_{1}
\ldots
\sum_{k_{N}}\int_{0}^{\infty}\D\omega_{N}\int \D^3r_{N}
\\ \times
C_{\lambda k_{1}\ldots\lambda k_{N}}
(\mathbf{r}_{1},\omega_{1},\cdots,\mathbf{r}_{N},\omega_{N})
\\ \times
\hat{f}_{\lambda k_{N}}^{\dagger}(\mathbf{r}_{N},\omega_{N})\cdots
\hat{f}_{\lambda k_{1}}^{\dagger}(\mathbf{r}_{1},\omega_{1}).
\end{multline}
If the convention to use only
standardized wave functions is adopted, one may write
\begin{multline}
\label{C44}
F_{N}[\hat{f}_{\lambda' k'}^{\dagger}(\mathbf{r}',\omega')
+
\epsilon \delta_{\lambda\lambda'}
(\tensor{P}_{\lambda})_{k k'}
(\mathbf{r},\mathbf{r}',\omega')
\delta(\omega-\omega')]
\\
=
F_{N}[\hat{f}_{\lambda' k'}^{\dagger}(\mathbf{r}',\omega')
+
\epsilon \delta_{\lambda\lambda'}
\delta_{kk'}
\delta (\mathbf{r}-\mathbf{r}')
\delta(\omega-\omega')],
\end{multline}
which means that the right-hand side of Eq.~(\ref{C42}) may
be evaluated, for this functional, just as an ordinary functional
derivative, i.e.,
\begin{multline}
\label{C45}
\left[
\hat{f}_{\lambda k}(\mathbf{r},\omega),
F_{N}[\hat{f}_{\lambda' k'}^{\dagger}(\mathbf{r}',\omega')]
\right]
=
\frac{\delta F_{N}[\hat{f}_{\lambda' k'}^{\dagger}(\mathbf{r}',\omega')]}
{\delta \hat{f}_{\lambda k}^{\dagger}(\mathbf{r},\omega)}.
\end{multline}
But since, due to the
definition of the reduced state space, only
commutators of the type (\ref{C45}) (for all $N$) are required,
and since Eq.~(\ref{C45}) can be obtained from Eq.~(\ref{eq8-16-1}) in the
same way that Eq.~(\ref{C42}) has been obtained from Eq.~(\ref{eq8-16}),
Eq.~(\ref{eq8-16-1}) is generally valid for the reduced state space.


\section{Derivation of Eq.~(\ref{eq500})}
\label{AppG}
For
notational convenience, let us write here
the integral equation (\ref{eq8-7})
in the compact operator form
\begin{equation}
\label{G1}
\mathcal{K}\mathcal{K}^{\dagger}=\mathcal{\sigma},
\end{equation}
with $\mathcal{K}$, and $\mathcal{\sigma}$
being, respectively, the
operators
associated (for chosen $\omega$)
with the integral kernels
$\tensor{K}(\mathbf{r},\mathbf{r}',\omega)$
and $\tensor{\sigma}(\mathbf{r},\mathbf{r}',\omega)$.
Accordingly, Eqs.~(\ref{NL21-1}) and (\ref{B3}) read
\mbox{$\mathcal{K}'$ $\!=$ $\!\mathcal{K}\mathcal{V}$} and
$\mathcal{V}^{\dagger}\mathcal{V}$ $\!=$
$\mathcal{V}\mathcal{V}^{\dagger}$  $\!=$ $\!\mathcal{I}$,
respectively
($\mathcal{I}$, unit operator). Assuming that the
(Hermitian and positive) operator $\mathcal{\sigma}$ takes the form
\begin{equation}
\label{G1-1}
\mathcal{\sigma}=
\mathcal{\sigma}_{0}
+
\epsilon \mathcal{\sigma}_{1},
\end{equation}
with $\epsilon$ being a small, real parameter,
we may try to find a solution to Eq.~(\ref{G1}) by the
perturbative ansatz
\begin{equation}
\label{G1-2}
\mathcal{K}'
=
\mathcal{K}'_{0}
+\epsilon
\mathcal{K}'_{1}
+\ldots,
\end{equation}
where
$\mathcal{K}_{0}'$
is a solution to Eq.~(\ref{G1}) for $\epsilon$ $\!=$ $\!0$.
Substituting Eqs.~(\ref{G1-1}) and (\ref{G1-2}) into
Eq.~(\ref{G1}),
we see that the first-order correction $\mathcal{K}'_{1}$
obeys the equation
\begin{equation}
\label{G2}
\mathcal{K}'_{0}\mathcal{K}_{1}^{\prime\dagger}
+
\mathcal{K}'_{1}\mathcal{K}_{0}^{\prime\dagger}
=\mathcal{\sigma}_{1},
\end{equation}
which determines the Hermitian part of
$\mathcal{K}{'}_{0}\mathcal{K}{'}_{1}^{\dagger}$
(recall that $\mathcal{\sigma}_{1}$ is Hermitian),
whereas the anti-Hermitian part is left undetermined.
The solution to Eq.~(\ref{G2}) may therefore be written as
\begin{equation}
\label{G3}
\mathcal{K}'_{1}
={\textstyle\frac{1}{2}}
(\mathcal{\sigma}_{1}
+\mathcal{A})
\mathcal{M}_{0},
\end{equation}
where
$\mathcal{M}_{0}$ $\!=$ $\!(\mathcal{K}_{0}^{\prime\dagger})^{-1}$,
and
$\mathcal{A}$ $\!=$ $\!-\mathcal{A}^{\dagger}$
is an arbitrary anti-Hermitian operator, which may be simply set to
zero. Note that the freedom to choose $\mathcal{A}$
corresponds to the freedom to choose $\mathcal{V}$.
We hence obtain to first order in
\mbox{$\mathcal{\sigma}$ $\!-$ $\!\mathcal{\sigma}_{0}$}
\begin{equation}
\label{G4}
\mathcal{K}'
=
\mathcal{K}'_0
+
{\textstyle\frac{1}{2}}
(\mathcal{\sigma}-\mathcal{\sigma}_{0})
\mathcal{M}_{0}.
\end{equation}
If $\mathcal{\sigma}_{0}$, $\mathcal{K}'_{0}$,
and $\mathcal{\sigma}$ are identified with
the operators associated with the kernels
(\ref{eq210}), (\ref{eq181}) and (\ref{eq490}),
respectively, Eq.~(\ref{G4}) is just the operator
equivalent of Eq.~(\ref{eq500})
[$\tensor{K}{^\prime}(\mathbf{r},\mathbf{r}',\omega)$
$\!\leftrightarrow$
$\!\tensor{K}(\mathbf{r},\mathbf{r}',\omega)$].
To calculate explicitly
the integral kernel
$\tensor{M}_{0}(\mathbf{r},\mathbf{r}',\omega)$
associated with
$\mathcal{M}_{0}$, we consider
the Fourier representation
\begin{equation}
\label{G5}
\tensor{K}{^\prime}_{\!\!\! 0}
(\mathbf{r},\mathbf{r}',\omega)
=
\frac{1}{(2\pi)^{3}}
\int \D^3k\,
\tensor{K}{^\prime}_{\!\!\! 0}
(\mathbf{k},\omega)
e^{i\mathbf{k}\cdot(\mathbf{r}-\mathbf{r}')},
\end{equation}
where, according to Eq.~(\ref{eq140}),
\begin{equation}
\label{G6}
\tensor{K}{^\prime}_{\!\!\! 0}
(\mathbf{k},\omega)
=
\sigma_{\parallel}^{1/2}(\omega)
\bigl[\tensor{I}\pm
\alpha(\omega)
\mathbf{k}\times\tensor{I}\bigr]
\end{equation}
[$\alpha(\omega)$ $\!=$ $\![\gamma(\omega)/
\sigma_{\parallel}(\omega)]^{1/2}$
$\!>$ $\!0$], which shows that the kernel
$\tensor{M}_{0}(\mathbf{r},\mathbf{r}',\omega)$
can be given by the Fourier integral
\begin{multline}
\label{G6-1}
\tensor{M}_{0}(\mathbf{r,r'},\omega)
=
\sigma_{\parallel}^{-1/2}(\omega)
\\\times
\int \frac{\D^3k}{(2\pi)^3}\,
\bigl[\tensor{I}\mp \alpha(\omega) \mathbf{k}\times\tensor{I}
\bigr]^{-1}
e^{i\mathbf{k}\cdot(\mathbf{r}-\mathbf{r}')},
\end{multline}
and it is not difficult to prove that
\begin{multline}
\label{G7}
\bigl[\tensor{I}\pm \alpha(\omega)
\mathbf{k}\times\tensor{I}\bigr]^{-1}
=
\tensor{I}\mp \alpha(\omega)\,
\frac{\mathbf{k}\times \tensor{I}}{1+\alpha(\omega)^2 k^2}
\\
+\alpha(\omega)^2\,
\frac{\mathbf{k}\times\tensor{I}\times\mathbf{k}}
{1+\alpha(\omega)^2 k^2}\,
.
\end{multline}
Introducing the function
\begin{align}
\label{G10}
m_{0}(\mathbf{r},\mathbf{r}',\omega)
&
=
\frac{1}{(2\pi)^3}\int \D^3k \,
\frac{e^{i\mathbf{k}\cdot(\mathbf{r}-\mathbf{r}')}}{\alpha(\omega)^{-2}+k^2}
\nonumber\\
&
=
-(4\pi|\mathbf{r}-\mathbf{r}'|)^{-1}
e^{-|\mathbf{r}-\mathbf{r}'|/\alpha(\omega)},
\end{align}
we may rewrite Eq.~(\ref{G6-1}) [with Eq.~(\ref{G7})]
to obtain Eq.~(\ref{G9}).
Note that the (Yukawa-type) function
$m_{0}(\mathbf{r},\mathbf{r}',\omega)$
satisfies the equation
\begin{equation}
\left[
- \Delta + \alpha(\omega)^{-2}
\right]
m_{0}(\mathbf{r},\mathbf{r}',\omega)
=\delta(\mathbf{r}-\mathbf{r}')
\end{equation}
together with the
boundary condition
$m_{0}(\mathbf{r},\mathbf{r}',\omega)
\to 0$
for
$|\mathbf{r}$ $\!-$ $\!\mathbf{r}'|\to\infty$.


%
\section{Bulk-medium Green tensor
and Kliever-Fuchs impedance}
\label{AppD}
For (translationally invariant) bulk material,
\begin{equation}
\label{D1}
\tensor{Q}(\mathbf{r},\mathbf{r}',\omega)
=
\int\frac{\D^3k}{(2\pi)^{3}}\,\,e^{i\mathbf{k}
\cdot(\mathbf{r-r'})}
\tensor{Q}(\mathbf{k},\omega),
\end{equation}
the solution to
Eq.~(\ref{eq8-3}) has the form
\begin{equation}
\label{D1-1}
\tensor{G}{^{(0)}}
(\mathbf{r},\mathbf{r}',\omega)
=
\int\frac{\D^3k}{(2\pi)^{3}}\,\,e^{i\mathbf{k}
\cdot(\mathbf{r-r'})}
\tensor{G}{^{(0)}}(\mathbf{k},\omega),
\end{equation}
where $\tensor{G}{^{(0)}}(\mathbf{k},\omega)$ is the solution to a
simple $3\times 3$ matrix equation.
In particular, for an isotropic medium without optical activity,
\begin{equation}
\label{D1-2}
\tensor{Q}(\mathbf{k},\omega)
=
Q_{\parallel}(k,\omega)
\frac{\mathbf{k}\ot\mathbf{k}}{k^2}
+
Q_{\perp}(k,\omega)
\left(\tensor{I}-\frac{\mathbf{k}\ot\mathbf{k}}{k^2}\right),
\end{equation}
one finds that
\begin{equation}
\label{D1-3}
\tensor{G}{^{(0)}}(\mathbf{k},\omega)
= \frac{\tensor{I}-\mathbf{k}\ot \mathbf{k}/k^2}
{D_{\perp}(k,\omega)}
-
\frac{\mathbf{k}\ot \mathbf{k}/k^2}
{D_{\parallel}(k,\omega)}
\,,
\end{equation}
where
\begin{align}
\label{C4-c}
&
D_{\perp}(k,\omega)
=
k^2-\omega^2/c^2-i\mu_{0}\omega
Q_{\perp}(k,\omega),
\\
\label{C5-a}
&
D_{\parallel}(k,\omega)
=
\omega^2/c^2+i\mu_{0}\omega
Q_{\parallel}(k,\omega),
\end{align}
and
\begin{equation}
\label{C10-a}
Q_{\parallel(\perp)}(k,\omega)
=
-i\varepsilon_{0}\omega
[\varepsilon_{\parallel(\perp)}(k,\omega)-1]
\end{equation}
in `dielectric' notation. In the general case
\begin{multline}
\label{D2}
\tensor{Q}(\mathbf{k},\omega)
=
Q_{\parallel}(\mathbf{k},\omega)\frac{\mathbf{k}\ot\mathbf{k}}{k^2}
\\
+
\left(\tensor{I}-\frac{\mathbf{k}\ot\mathbf{k}}{k^2}\right)
\cdot
\tensor{Q}_{\perp}(\mathbf{k},\omega)
\cdot
\left(\tensor{I}-\frac{\mathbf{k}\ot\mathbf{k}}{k^2}\right),
\end{multline}
$D_{\parallel}(k,\omega)$ changes to
$D_{\parallel}(\mathbf{k},\omega)$,
$D_{\perp}(k,\omega)$ changes to the tensor
\begin{equation}
\label{C4-a}
\tensor{D}_{\perp}(\mathbf{k},\omega)
=
k^2-\omega^2/c^2-i\mu_{0}\omega \tensor{Q}_{\perp}(\mathbf{k},\omega),
\end{equation}
and the first term
on the right-hand side of Eq.~(\ref{D1-3})
has to be replaced
according to
($\tensor{I}_{k}$ $\!=$ $\!\tensor{I}$ $\!-$
$\!\mathbf{k}\ot\mathbf{k}/{k^2}$)
\begin{equation}
\label{C4-a-1}
\frac{\tensor{I}-\mathbf{k}\ot\mathbf{k}/k^2}
%
{D_{\perp}(k,\omega)}
\mapsto
\tensor{I}_{k}
\cdot
\left[\tensor{I}_{k}
\cdot
\tensor{D}_{\perp}(\mathbf{k},\omega)
\cdot
\tensor{I}_{k}
\right]^{-1}
\!\!\cdot
\tensor{I}_{k}.
\end{equation}

It may be convenient---particularly with
regard to systems that are translationally
invariant only in a plane, say the
$xy$~plane---to rewrite Eq.~(\ref{D1-1}) as
[$\mathbf{r}$ $\!=$ $\!(\vec{\rho},z)$,
$\mathbf{k}$ $\!=$ $\!(\mathbf{q,\beta})$]
\begin{equation}
\label{C3a}
\tensor{G}{^{(0)}}(\mathbf{r},\mathbf{r}',\omega)
=
\int \D^2q \,e^{i\mathbf{q}\cdot(\mathbf{\vec{\rho}-\vec{\rho'}})}
\tensor{G}{^{(0)}}(
z,z',
\mathbf{q},\omega),
\end{equation}
where
\begin{equation}
\label{C3c}
\tensor{G}{^{(0)}}(z,z',\mathbf{q},\omega)
=\int\frac{\D\beta}{(2\pi)^{3}}\,\,e^{i\beta(z-z')}
\tensor{G}{^{(0)}}(\mathbf{k},\omega).
\end{equation}
The analytical
properties of the integrand in Eq.~(\ref{C3c})
with respect to $\beta$ depend
on the specific decay to zero
of $\tensor{Q}(\mathbf{r-r'},\omega)$ for $|\mathbf{r-r'}|
\to\infty$.
For sufficiently rapid decay,
Eq.~(\ref{C3c}) admits an evaluation by contour integration in the
complex $\beta$~plane, which
will be governed---focusing again on isotropic media without
optical activity---by the solutions
$\beta$ $\!=$ $\!\beta_{\nu}^{\perp,\parallel}(q,\omega)$
of the two dispersion
equations $D_{\perp,\parallel}(k,\omega)$ $\!=$ $\!0$.
In contrast
to the case of
spatially non-dispersive material,
these equations are transcendental with respect
to $\beta$ rather than polynomial, so that nothing
can be said about the number of their solutions
in general.
Specifically, if there are more than
two functions
$\beta_{\nu}^{\perp}(q,\omega)$ [and/or one or more functions
$\beta_{\nu}^{\parallel}(q,\omega)$], the medium
is said to support `additional' (inhomogeneous plane-)waves.

We close this appendix with the following
(perhaps not well-known) observation.
Inserting Eq.~(\ref{D1-3}) in Eq.~(\ref{C3c}) and
setting $z\!-\!z'\!=\!0\pm$,
and making use of the decompositions
\begin{align}
\label{E13}
&\tensor{I}-\mathbf{k}\ot\mathbf{k}/k^2=\uv{s}(\mathbf{q})\ot\uv{s}(\mathbf{q})
+\uv{p}(\mathbf{k})\ot\uv{p}(\mathbf{k})
\\
\label{E14}
&\tensor{I}_{z}=\uv{s}(\mathbf{q})\ot\uv{s}(\mathbf{q})
+\mathbf{q}\ot\mathbf{q}/q^2,
\end{align}
with $\uv{s}(\mathbf{q})$ $\!=$ $\!\mathbf{q}\times\uv{z}/q$ and
$\uv{p}(\mathbf{k})$ $\!=$ $\!\mathbf{k}\times\mathbf{q}\times\uv{z}/kq$
being polarization unit vectors, one can
show that
\begin{multline}
\label{E15}
\tensor{I}_{z}\cdot\tensor{G}{^{(0)}}(
0\pm,0,
\mathbf{q},\omega)\cdot
\tensor{I}_{z}
\\
=
(i\mu_{0}\omega)^{-1}
\left[
{Z}_{s}(q,\omega)\,
\uv{s}(\mathbf{q})\uv{s}(\mathbf{q})+
{Z}_{p}(q,\omega)\,
\mathbf{qq}/q^2
\right],
\end{multline}
where
\begin{align}
\label{C11-a}
&{Z}_{s}(q,\omega)
=
i\mu_{0}\omega\int_{-\infty}^{\infty}\frac{\D\beta}{(2\pi)^3}
\frac{e^{i\beta 0 \pm}}
{D_{\perp}(k,\omega)}\,,
\\
\label{C12-a}
&{Z}_{p}(q,\omega)
=i\mu_{0}\omega
\nonumber\\&\hspace{2ex}\times
\int_{-\infty}^{\infty}
\frac{\D\beta}{(2\pi)^3}\frac{e^{i\beta 0\pm}}{k^2}
\left[\frac{\beta^2}
{D_{\perp}(k,\omega)}-\frac{q^2}
{D_{\parallel}(k,\omega)}
\right].
\end{align}
With Eqs.~(\ref{C4-c})--(\ref{C10-a}),
Eqs.~(\ref{C11-a}) and (\ref{C12-a})
are recognized (up to a trivial factor)
as the surface impedance expressions first
derived by Kliever and Fuchs
\cite{KlieverKL081968} for a
spatially dispersive half-space by assuming
specular electron reflection
(see also Ref.~\cite{EsquivelR062004}).


\section{Derivation of Eqs.~(\ref{NL44a}) and (\ref{NL44b})}
\label{AppE}

From inspection of Eq.~(\ref{NL42}) it is seen that only
the tangential components of the electric field,
which are assumed to be continuous
at $z'\!=\!0$ and $z'\!=\!d$, contribute to
the right-hand side of this equation.
We therefore evaluate
Eq.~(\ref{NL42}) at $z'$ $\!=$ $\!0+$ and $z'$ $\!=$ $\!d-$
and take the tangential ($\tensor{I}_{z}$)
component thereof to obtain the
linear equations (the arguments $\mathbf{q}$ and $\omega$ are kept
fixed in this appendix and are suppressed for
notational convenience)
\begin{multline}
\label{B1-a}
\fon{E}(0+)\cdot
\tensor{I}_{z}\cdot
[\tensor{I}+\tensor{R}(0,0+)]
\cdot\tensor{I}_{z}
\\
-\fon{E}(d-)\cdot
\tensor{I}_{z}
\cdot
\tensor{R}(d,0+)
\cdot
\tensor{I}_{z}
=\fon{E}^{\mathrm{(in)}}(0+)
\cdot
\tensor{I}_{z}
\end{multline}
and
\begin{multline}
\label{B2-a}
\fon{E}(0+)\cdot
\tensor{I}_{z}
\cdot
\tensor{R}(0,d-)
\cdot
\tensor{I}_{z}
\\
+\fon{E}(d-)\cdot
\tensor{I}_{z}
\cdot
[\tensor{I}-\tensor{R}(d,d-)]
\cdot
\tensor{I}_{z}
=\fon{E}^{\mathrm{(in)}}(d-)\cdot
\tensor{I}_{z},
\end{multline}
respectively,
which are to be solved
for the tangential components
$\fon{E}(0+)\cdot\tensor{I}_{z}$
and $\fon{E}(d-)\cdot\tensor{I}_{z}$.
To represent the solution in a compact form, we assign
to any matrix $\tensor{A}$
satisfying
$\tensor{I}_{z}\cdot\tensor{A}\cdot\tensor{I}_{z}
\!=\!\tensor{A}$
its inverse on the $\tensor{I}_{z}$-space,
$\tensor{A}{^{\sharp}}
\!=\!\tensor{I}_{z}\cdot(\tensor{I}_{z}\cdot\tensor{A}
\cdot\tensor{I}_{z})^{-1}\cdot\tensor{I}_{z}$.
It is straightforward to see
that the block-matrix formula
\begin{multline}
\label{E3}
{\begin{pmatrix}
\tensor{A} & \tensor{B} \\
\tensor{C} & \tensor{D}
\end{pmatrix}}
\cdot
\begin{pmatrix}
\tensor{C}{^{\sharp}}
\!\cdot\!
[\tensor{A}\!\cdot\!\tensor{C}{^{\sharp}}
\!-\!\tensor{B}\!\cdot\!\tensor{D}{^{\sharp}}]^{\sharp}
&
\tensor{A}{^{\sharp}}\!\cdot\![\tensor{C}\!\cdot\!\tensor{A}{^{\sharp}}
\!-\!\tensor{D}\!\cdot\!\tensor{B}{^{\sharp}}]{^{\sharp}} \\
\tensor{D}{^{\sharp}}\!\cdot\![\tensor{B}\!\cdot\!\tensor{D}{^{\sharp}}
\!-\!\tensor{A}\!\cdot\!\tensor{C}{^{\sharp}}]^{\sharp}
&
\tensor{B}{^{\sharp}}\!\cdot\![\tensor{D}\!\cdot\!\tensor{B}{^{\sharp}}
\!-\!
\tensor{C}\!\cdot\!\tensor{A}{^{\sharp}}
]^{\sharp}
\end{pmatrix}
\\
=
\begin{pmatrix}
\tensor{I}_{z} & 0 \\
0 & \tensor{I}_{z}
\end{pmatrix}
\end{multline}
is generally valid whenever the requisite inverse elements exist,
so that the solution to Eqs.~(\ref{B1-a}) and (\ref{B2-a})
can be written in the form
\begin{align}
\label{E4}
&\fon{E}(0+)\cdot \tensor{I}_{z}
=
[\fon{E}^{\mathrm{(in)}}(0+)\cdot\tensor{C}{^{\sharp}}
\nonumber\\&\hspace{14ex}
-\fon{E}^{\mathrm{(in)}}(d-)\cdot\tensor{D}{^{\sharp}}
]\cdot[\tensor{A}\cdot\tensor{C}{^{\sharp}}
-\tensor{B}\cdot\tensor{D}{^{\sharp}}]^{\sharp},
\\
\label{E5}
&\fon{E}(d-)\cdot \tensor{I}_{z}
=
[
\fon{E}^{\mathrm{(in)}}(0+)\cdot\tensor{A}{^{\sharp}}
\nonumber\\&\hspace{14ex}
-
\fon{E}^{\mathrm{(in)}}(d-)\cdot\tensor{B}{^{\sharp}}
]\cdot
[\tensor{C}\cdot\tensor{A}{^{\sharp}}
-\tensor{D}\cdot\tensor{B}{^{\sharp}}]^{\sharp}
\end{align}
together with Eqs.~(\ref{NL50-a})--(\ref{NL50-d}).
Recalling again the continuity of the tangential
component of the electric field, we are left with
Eq.~(\ref{NL44a}) and (\ref{NL44b}).
It should be noted that
the above inversion procedure
fails at particular values of
the (suppressed) arguments $\mathbf{q}$ and
$\omega$, because of singularities.
However, as we are dealing with a
lossy system, such singularities---corresponding to
guided waves---may appear only when $\Im \omega\!<\!0$
(for real values of $\mathbf{q}$).
%


\section{Derivation of Eqs.~(\ref{NL53-a})--(\ref{NL53-d})}
\label{AppF}

Let us attribute to the three regions
$j$ $\!=$ $\!0$ ($z$ $\!<$ $\!0$), $j$ $\!=$ $\!1$ ($0$ $\!<$ $\!z$ $\!<$
$\!d$), \mbox{$j$ $\!=$ $\!2$} ($z$ $\!>$ $\!d$)
bulk-medium conductivities
$\tensor{Q}_{j}(\mathbf{r}$ $\!-$ $\mathbf{r}',\omega)$,
which combine to the
overall conductivity tensor in the sense of
Eq.~(\ref{eq200})
[$j\leftrightarrow\mathbf{L}$].
For each of these regions, we construct,
according to Eq.~(\ref{NL33}),
the translationally invariant
bulk-medium Green tensor
\mbox{$\tensor{G}_{j}(\mathbf{r},\mathbf{r}',\omega)$}
and the associated auxiliary tensor
\mbox{$\tensor{\Gamma}_{j}(\mathbf{r},\mathbf{r}',\omega)$}
in terms of their Fourier components
$\tensor{G}_{j}(z,z',\mathbf{q},\omega)$ and
$\tensor{\Gamma}_{j}(z,z',\mathbf{q},\omega)$
[defined according to Eqs.~(\ref{C3a}), (\ref{C3c})].
The three regions
are thus described on an equal footing so that
for the field in each region,
an equation similar to Eq.~(\ref{NL210}) holds.
Evaluating the tangential components of
these three equations
(which together determine the field in all of
space) and using
the continuity conditions
at $z$ $\!=$ $\!0$ and $z$ $\!=$ $\!d$,
one obtains two sets of equations for
the tangential boundary values,
\begin{widetext}
\begin{multline}
\label{F1}
\bigl\{
[\fon{E}(d)\times\uv{z}] \cdot
\tensor{\Gamma}_{1}(d,d-)
+
[
\mu_{0}^{-1}
\fon{B}(d)\times\uv{z}] \cdot
\tensor{G}_{1}(d,d-)
-
[\fon{E}(0)\times\uv{z}] \cdot
\tensor{\Gamma}_{1}(0,d-)
-
[
\mu_{0}^{-1}
\fon{B}(0)\times\uv{z}] \cdot
\tensor{G}_{1}(0,d-)
\bigr\}\cdot\tensor{I}_{z}
\\
=
-
\bigl\{
[\fon{E}(d)\times\uv{z}] \cdot
\tensor{\Gamma}_{2}(0,0+)
+
[
\mu_{0}^{-1}
\fon{B}(d)\times\uv{z}] \cdot
\tensor{G}_{2}(0,0+)
\bigr\}\cdot\tensor{I}_{z}
\end{multline}
and
\begin{multline}
\label{F2}
\bigl\{
[\fon{E}(d)\times\uv{z}] \cdot
\tensor{\Gamma}_{1}(d,0+)
+
[
\mu_{0}^{-1}
\fon{B}(d)\times\uv{z}] \cdot
\tensor{G}_{1}(d,0+)
-
[\fon{E}(0)\times\uv{z}] \cdot
\tensor{\Gamma}_{1}(0,0+)
-
[
\mu_{0}^{-1}
\fon{B}(0)\times\uv{z}] \cdot
\tensor{G}_{1}(0,0+)
\bigr\}\cdot\tensor{I}_{z}
\\
=
\bigl\{
[\fon{E}(0)\times\uv{z}] \cdot
\tensor{\Gamma}_{0}(0,0-)
+
[
\mu_{0}^{-1}
\fon{B}(0)\times\uv{z}] \cdot
\tensor{G}_{0}(0,0-)
\bigr\}\cdot\tensor{I}_{z},
\end{multline}
\end{widetext}
where,
for notational convenience,
the arguments $\mathbf{q}$ and $\omega$ of
$\fon{E}$ and
$\fon{B}$,
and the arguments $-\mathbf{q}$ and $\omega$ of $\tensor{G}_{j}$ and
$\tensor{\Gamma}_{j}$ have been suppressed.
Solving these linear relations for
$\fon{B}(0)\times\uv{z}$ and
$\fon{B}(d)\times\uv{z}$ in terms of
$\fon{E}(0)\times\uv{z}$ and
$\fon{E}(d)\times\uv{z}$,
we need only compare the result with
Eq.~(\ref{NL40}) to
verify
Eqs.~(\ref{NL53-a})--(\ref{NL53-d}).
This can be conveniently
done by representing Eqs.~(\ref{F1}) and (\ref{F2})
in the form
\begin{equation}
\label{F3}
%
\mu^{-1}\!
\begin{pmatrix}
\fon{B}(0)\times\uv{z} \\
\fon{B}(d)\times\uv{z}
\end{pmatrix}
^{\!\mathsf{T}}\!
\cdot
\begin{pmatrix}
\tensor{M} & \tensor{N} \\
\tensor{P} & \tensor{S}
\end{pmatrix}
=
\begin{pmatrix}
\fon{E}(0)\times\uv{z} \\
\fon{E}(d)\times\uv{z}
\end{pmatrix}
^{\!\mathsf{T}}\!
\cdot
\begin{pmatrix}
\tensor{T} & \tensor{U} \\
\tensor{V} & \tensor{W}
\end{pmatrix},
\end{equation}
where $\tensor{M},\tensor{N},\tensor{P},\tensor{S},
\tensor{T},\tensor{U},\tensor{V},\tensor{W}$
are given in Eqs.~(\ref{NL51-a})--(\ref{NL52-d}).
Rewriting Eq.~(\ref{NL40}) ($z$ $\!=$ $\!0,d$) in
an analogous form and applying to Eq.~(\ref{F3})
the inversion formula (\ref{E3}),
one obtains
\begin{multline}
\label{F4}
%
\begin{pmatrix}
-\tensor{Y}(0,0) & -\tensor{Y}(0,d) \\
\tensor{Y}(d,0) & \tensor{Y}(d,d)
\end{pmatrix}
=
\frac{1}{(2\pi)^{2}}
\begin{pmatrix}
\tensor{T} & \tensor{U} \\
\tensor{V} & \tensor{W}
\end{pmatrix}
\cdot
\\\times
\begin{pmatrix}
\tensor{P}{^{\sharp}}\!\cdot\!
[\tensor{M}\!\cdot\!\tensor{P}{^{\sharp}}
\!-\!\tensor{N}\!\cdot\!\tensor{S}{^{\sharp}}]^{\sharp}
&
\tensor{M}{^{\sharp}}\!\cdot\![\tensor{P}\!\cdot\!\tensor{M}{^{\sharp}}
\!-\!\tensor{S}\!\cdot\!\tensor{N}{^{\sharp}}]^{\sharp} \\
\tensor{S}{^{\sharp}}\!\cdot\![\tensor{N}\!\cdot\!\tensor{S}{^{\sharp}}
\!-\!\tensor{M}\!\cdot\!\tensor{P}{^{\sharp}}]^{\sharp}
&
\tensor{N}{^{\sharp}}\!\cdot\![\tensor{S}\!\cdot\!\tensor{N}{^{\sharp}}
\!-\!
\tensor{P}\!\cdot\!\tensor{M}{^{\sharp}}
]^{\sharp}
\end{pmatrix},
\end{multline}
which immediately leads to
Eqs.~(\ref{NL53-a})--(\ref{NL53-d}).
\end{appendix}
\bibliographystyle{apsrev}
\bibliography{QuantNLBib5}
\end{document}